\documentclass[prd,reprint,superscriptaddress,showpacs,preprintnumbers,nobibnotes,amsmath,amssymb,aps,longbibliography]{revtex4-2}

\usepackage{cancel}
\usepackage{accents}
\usepackage{mciteplus,slashed}
\usepackage{amssymb,cancel,amsmath}
\usepackage{dcolumn}
\usepackage{bm}
\usepackage[caption=false]{subfig}
\usepackage{appendix}
\usepackage{feynmp-auto}
\usepackage[T1]{fontenc}
\usepackage{csvsimple}
\usepackage[colorlinks=true,citecolor=blue,linkcolor=blue]{hyperref}
\usepackage[section]{placeins}
\usepackage[capitalise]{cleveref}
\usepackage{booktabs}
\usepackage{graphicx}
\usepackage{subfig}
\usepackage{mathrsfs}
\usepackage{multirow}
\usepackage[utf8]{inputenc}
\usepackage{siunitx}
\usepackage{comment}
\usepackage{gensymb}
\usepackage{enumitem}

\usepackage[dvipsnames]{xcolor}
\usepackage[normalem]{ulem}

\newcommand{\gev}{\rm GeV}
\newcommand{\tev}{\rm TeV}
\newcommand{\UNsq}[1]{\ensuremath{|U_{#1 N}|^2}}

\begin{document}

\title{Novel Heavy Neutral Lepton Searches With Next-Generation Muon Accelerators}

\author{Carlos A. Arg\"uelles}
\affiliation{Department of Physics \& Laboratory for Particle Physics and Cosmology, Harvard University, Cambridge, MA 02138, USA\looseness=-1}
\author{Gonzalo Herrera}
\affiliation{Kavli Institute for Astrophysics and Space Research, Massachusetts Institute of Technology, Cambridge, MA 02139, USA\looseness=-1}
\affiliation{Department of Physics \& Laboratory for Particle Physics and Cosmology, Harvard University, Cambridge, MA 02138, USA\looseness=-1}
\author{Nicholas Kamp}
\email{nkamp@fas.harvard.edu}
\affiliation{Department of Physics \& Laboratory for Particle Physics and Cosmology, Harvard University, Cambridge, MA 02138, USA\looseness=-1}
\author{Peng-Cheng Lu}
\email{pclu@sdu.edu.cn}
\affiliation{School of Physics, Shandong University, Jinan, Shandong 250100, China\looseness=-1}
\affiliation{Department of Physics and Astronomy, Michigan State University, East Lansing, MI 48824, USA}
\author{Keping Xie}
\email{kepingxie@sjtu.edu.cn}
\affiliation{State Key Laboratory of Dark Matter Physics \& Key Laboratory for Particle Astrophysics and Cosmology (MOE) \& Shanghai Key Laboratory for Particle Physics and Cosmology, Tsung-Dao Lee Institute \& School of Physics and Astronomy, Shanghai Jiao Tong University, Shanghai 201210, China\looseness=-1}
\affiliation{Department of Physics and Astronomy, Michigan State University, East Lansing, MI 48824, USA}

\begin{abstract}
We investigate the production and detection of heavy neutral leptons (HNLs) at a next-generation muon facility, where an intense multi-\tev\ muon beam is available either as a stored beam or as a collider beam.
Three proposals are considered: (1) a heavy neutrino factory, with HNL production in muon decays in a storage ring and detection in a long, low-density detector aligned with a straight section;
(2) a muon beam dump, studied with a dense tungsten or water target followed by a helium-filled or evacuated decay volume, and with an Earth target paired with either an atmospheric decay region observed by an air-Cherenkov camera or an evacuated tunnel;
and (3) a muon-ion collider, in which HNLs are reconstructed as displaced vertices.
Assuming $N_\mu = 10^{22}$ muons, the heavy neutrino factory improves upon existing $\UNsq{\mu}$ constraints below the muon mass by two orders of magnitude.
The muon beam dump configurations would improve upon existing LHC constraints at $m_N\sim 10~{\rm GeV}$ by up to six orders of magnitude, and improve upon projected high-luminosity LHC sensitivities by two orders of magnitude.
Furthermore, the muon beam dump HNL searches improve upon electron mode Future Circular Collider sensitivities by at least one order of magnitude in the same mass regime.
The muon-ion collider probes HNL masses up to $m_N \sim 100~\gev$ with sensitivity comparable to the electron mode of the Future Circular Collider.
Taken together, the projections cover $2 m_e \lesssim m_N \lesssim 100~\gev$ without gaps.
The beam dump proposals in particular would cover a substantial fraction of the parameter space compatible with Akhmedov--Rubakov--Smirnov leptogenesis, and would be the first experiments to reach the type-I seesaw floor for $m_N > 1~\gev$.


\end{abstract}

\maketitle
\tableofcontents

\section{Introduction}

\begin{figure*}
    \centering
    \includegraphics[width=1.05\linewidth]{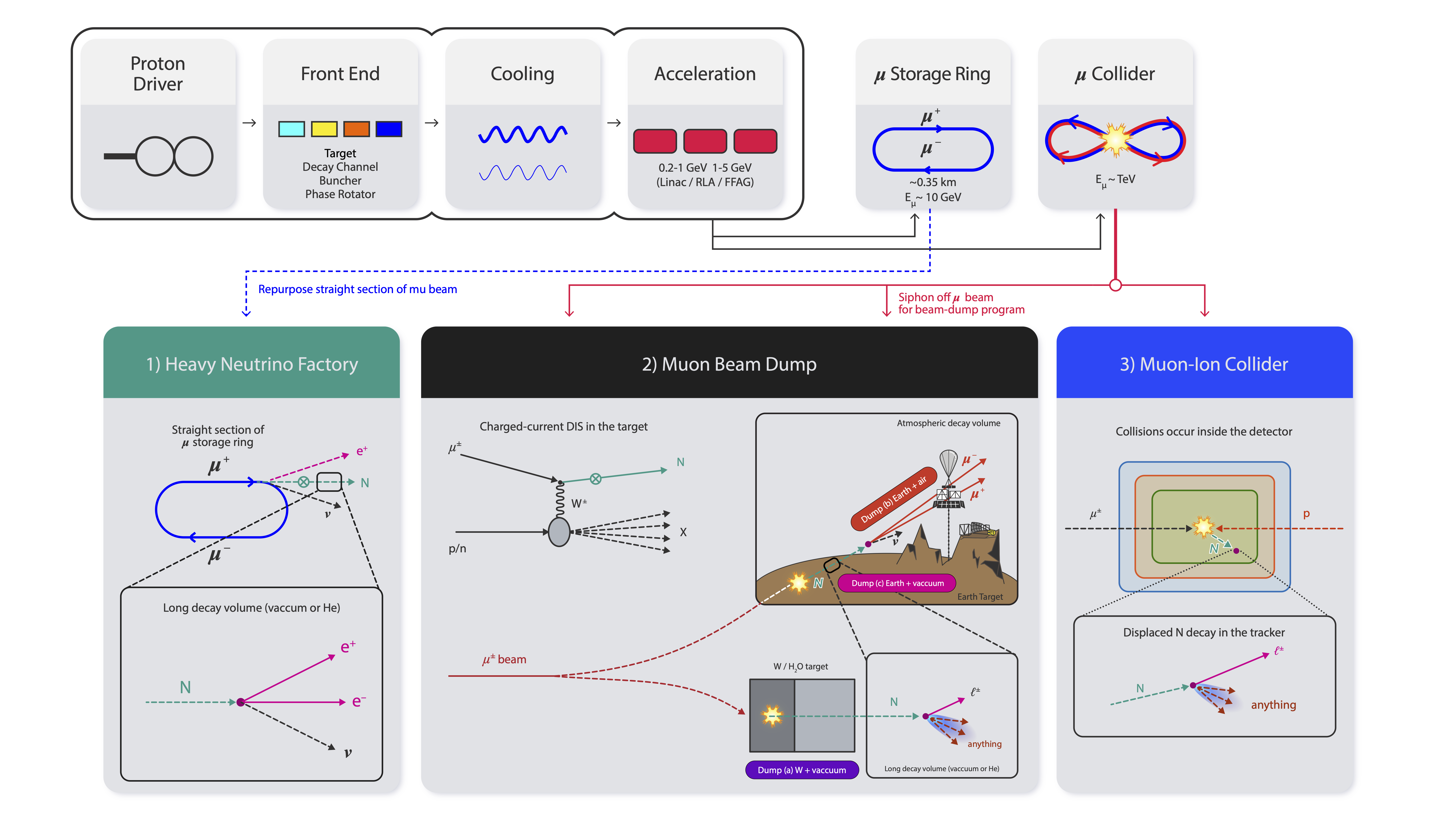}
    \caption{Schematic of the three proposals: a heavy neutrino factory, a muon beam dump with dense-target and atmospheric configurations, and a muon-ion collider. The Earth-plus-vacuum beam-dump configuration combines the Earth target with an evacuated decay tunnel.}
    \label{fig:diagram}
\end{figure*}

Among the most important open questions in particle physics are the origin of neutrino masses and the baryon asymmetry of the Universe.
An elegant solution to both problems is found in the form of heavy neutral leptons (HNLs), right-handed counterparts to the left-handed active neutrinos~\cite{Abdullahi:2022jlv}.
Such HNLs can famously generate neutrino masses via the seesaw mechanism~\cite{Minkowski:1977sc,Mohapatra:1979ia,Gell-Mann:1979vob,Yanagida:1979as}
and seed the baryon asymmetry of the early Universe through CP-violating oscillations~\cite{Akhmedov:1998qx,Asaka:2005pn}, later converted by sphalerons~\cite{Kuzmin:1985mm,Fukugita:1986hr}.

Despite its attractiveness, the HNL solution is difficult to test experimentally.
This is because HNL models that explain the active neutrino masses tend to have masses that are too large and/or couplings that are too small to probe in any existing laboratory environment.
Intriguingly, the first problem can be remedied in models of low-scale seesaw and leptogenesis, in which the HNL mass can be brought down to the TeV scale or lower~\cite{Gorbunov:2007ak,Atre:2009rg,Ibarra:2011xn,Drewes:2016jae}.
However, even in these scenarios the typical couplings are out of reach of current experimental searches~\cite{Drewes:2019mhg}.

At the same time, a paradigm shift is on the horizon of experimental high energy physics in the form of a muon collider~\cite{Accettura:2023ked,InternationalMuonCollider:2025sys,Begel:2025ldu}.
The muon collider is a proposed next-generation collider that would reach energy scales an order of magnitude beyond the Large Hadron Collider~\cite{AlAli:2021let}.
At the same time, the high-intensity TeV muon beam will unavoidably decay into a high-intensity TeV neutrino beam~\cite{Geer:1997iz,Bojorquez-Lopez:2024bsr}.
Previous works have studied the rich physics program enabled by this neutrino beam, including precision electroweak physics~\cite{Mangano:2001mj,deGouvea:2025zfq,Marzocca:2025inb}, neutrino cross sections~\cite{Harris:1997xj,King:1999kx,nuSTORM:2013cqr,Burk:2026fox,Choi:2026yzw}, QCD~\cite{Mangano:2001mj,Forte:2001ph}, and long-baseline oscillations~\cite{Denton:2024glz,Kamp:2025yzq}.

In this work, we consider this muon beam as a means to produce a high-intensity source of HNLs.
Previous work~\cite{Choi:2026yzw} has considered the associated neutrino beam as a production source for HNL searches.
Here we focus on HNL production by the muons themselves, via either decay or scattering.
These HNLs are then detected via their decays to unique final states, often enabling background-free searches.
We explore three novel experimental proposals leveraging this idea:
\begin{itemize}
    \item \textbf{``Heavy'' neutrino factory:} HNL production in muon decays in a storage ring and detection in a long, low-density detector
    \item \textbf{Muon beam dump:} HNL production from muon scattering, with three configurations pairing a dense target with a low-density or evacuated decay volume, the Earth with an atmospheric decay region observed by an air Cherenkov camera, and the Earth with an evacuated decay tunnel
    \item \textbf{Muon-ion collider:} HNL production in muon-ion collisions and detection in an LHC-style detector at the interaction point
\end{itemize}
The three proposals and the beam-dump building blocks are depicted in \cref{fig:diagram}.
In combination, they expand the sensitivity to HNLs by many orders of magnitude across a wide range of HNL masses up to $m_N \sim 100~{\rm GeV}$, covering a region of the parameter space motivated by ARS (Akhmedov--Rubakov--Smirnov) leptogenesis~\cite{Akhmedov:1998qx}.
Notably, the tungsten-plus-vacuum and Earth-plus-vacuum beam-dump configurations provide the first proposed experimental sensitivities to seesaw-motivated HNLs for $m_N > 1~{\rm GeV}$.

The rest of this article discusses the three HNL search proposals in more detail: \cref{sec:decay} covers the heavy neutrino factory, \cref{sec:beam_dump} the muon beam dump, and \cref{sec:muon_ion} the muon-ion collider. Within the beam-dump proposal, \cref{sec:dump} treats dense targets with low-density or evacuated decay volumes, \cref{sec:balloon} an Earth target with an atmospheric decay region, and \cref{sec:terrestrial_vacuum} an Earth target with an evacuated decay tunnel.
\Cref{sec:leptogenesis} then discusses the leptogenesis models motivating the region of parameter space covered by our proposals.
Finally, \cref{sec:combined} combines the projections and concludes.



\section{The Heavy Neutrino Factory} \label{sec:decay}

\begin{figure}
    \centering
    \includegraphics[width=0.42\linewidth]{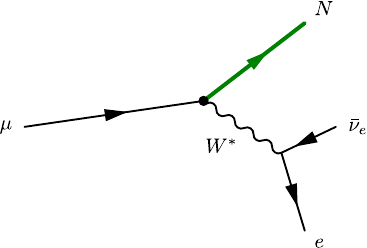}
    \includegraphics[width=0.42\linewidth]{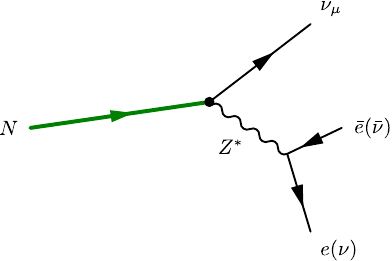}
    \caption{HNL production through the $\mu$ decay and its sub-sequential decay, through the $|U_{\mu N}|^2$ mixing.}
    \label{fig:mu_decays}
\end{figure}

Heavy neutral leptons with $\UNsq{\mu} > 0$ can be produced in muon decays via the diagrams in \cref{fig:mu_decays} and subsequently decay through the channel $N \to e^+ e^- \nu$.
This strategy was employed in Ref.~\cite{Ema:2023buz} to set strong constraints on HNLs for $m_N \lesssim m_\mu$ with data from the LSND experiment.
In this work, we investigate the potential of a similar search at a neutrino factory, which will produce orders of magnitude more muons than the LSND setup.

The HNL energy spectrum follows from the differential decay width for $\mu \to N e \nu$, which receives contributions from both the muon and electron mixing,
\begin{align}
    \frac{d\Gamma(\mu \to N e\nu)}{dE_N}
    &= \frac{G_F^2\, p_N}{12\pi^3} \Big\{ \UNsq{\mu}\big[3 E_N (m_\mu^2 + m_N^2) \nonumber\\
    &\qquad - 4 m_\mu E_N^2 - 2 m_\mu m_N^2\big] \nonumber\\
    &\quad + 6\,\UNsq{e}\, E_N\big[m_\mu^2 + m_N^2 - 2 m_\mu E_N\big]\Big\},
    \label{eq:mu_diff_width}
\end{align}
where $p_N = \sqrt{E_N^2 - m_N^2}$ and the HNL energy in the muon rest frame ranges over $m_N \le E_N \le (m_\mu^2 + m_N^2)/2m_\mu$.
The number of HNL decays to $e^+ e^-$ pairs in a detector of length $L_{\rm det}$ is given by~\cite{Ema:2023buz}
\begin{equation}
N_{ee} = N_\mu\, \epsilon_{\rm det}\,  \frac{1}{\Gamma_\mu} \int dE_N\, \frac{d\Gamma (\mu \to Ne\nu)}{dE_N}\, \frac{L_{\rm det} }{\gamma \beta c\, \tau_{N \to e^+ e^- \nu}},
\end{equation}
where $N_\mu$ is the number of muon decays, $\epsilon_{\rm det}$ is the probability that the resulting HNL passes through the detector, $\Gamma_\mu$ is the total muon decay width, $\gamma\beta$ is the HNL boost in the lab frame, and $\tau_{N \to e^+ e^- \nu}$ is the HNL lifetime against decay to $e^+ e^- \nu$.
The latter is set by the partial width
\begin{equation}\label{eq:Nee_width}
    \Gamma_{N \to e^+ e^- \nu} = \frac{G_F^2 m_N^5}{768\pi^3}\left[\UNsq{\mu}\,c_\mu + \UNsq{e}\,c_e\right],
\end{equation}
with $c_{\mu(e)} = 1 \mp 4\sin^2\theta_W + 8\sin^4\theta_W$. The electron-mixing channel receives an additional charged-current contribution that enhances $c_e$ relative to $c_\mu$.
We assume $N_\mu = 10^{22}$ muon decays~\cite{Denton:2024glz} and a detector of length $L_{\rm det} = 100~{\rm m}$ as a baseline.

\begin{figure}[h]
    \centering
    \includegraphics[width=1\linewidth]{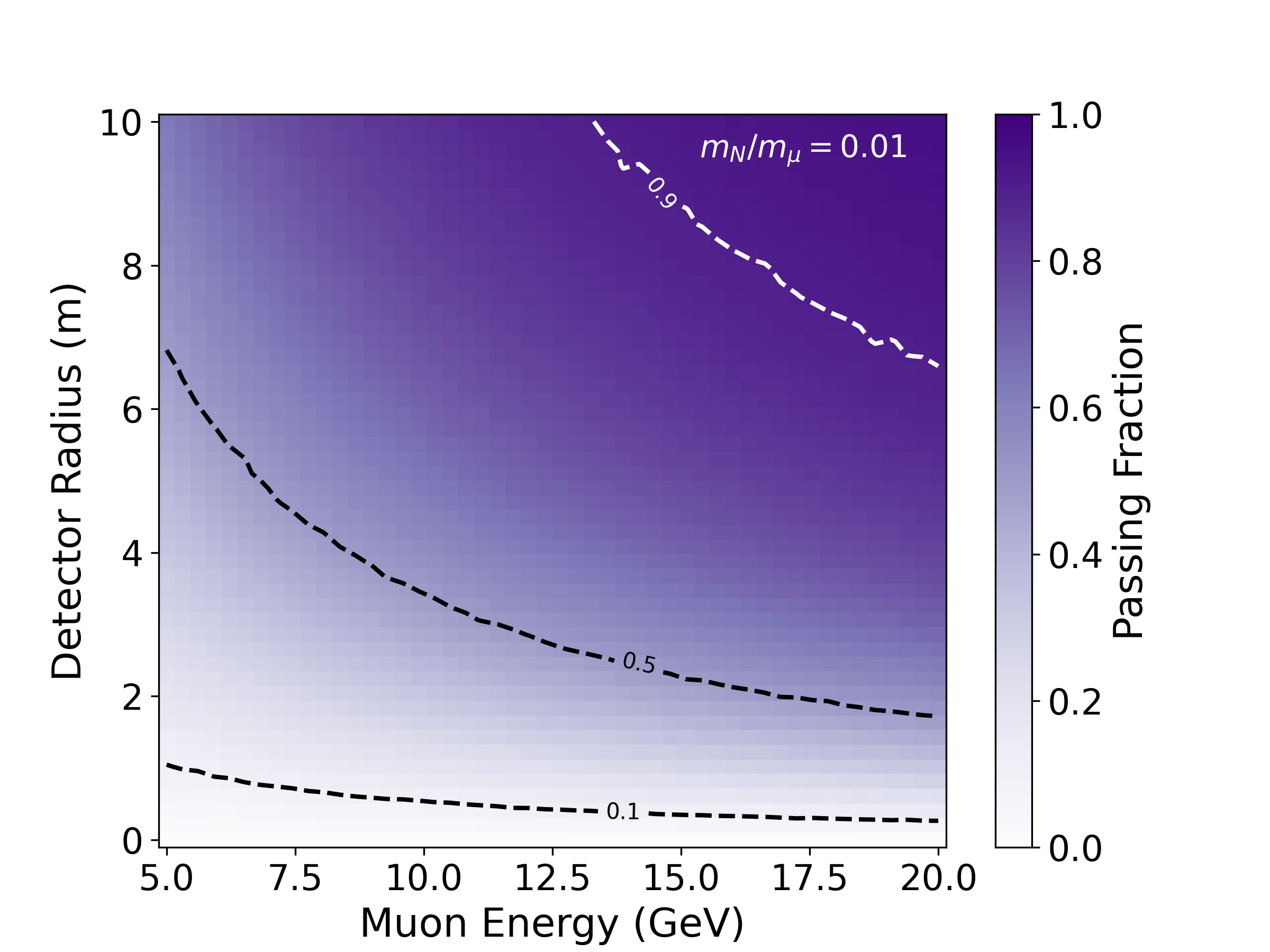}
    \includegraphics[width=1\linewidth]{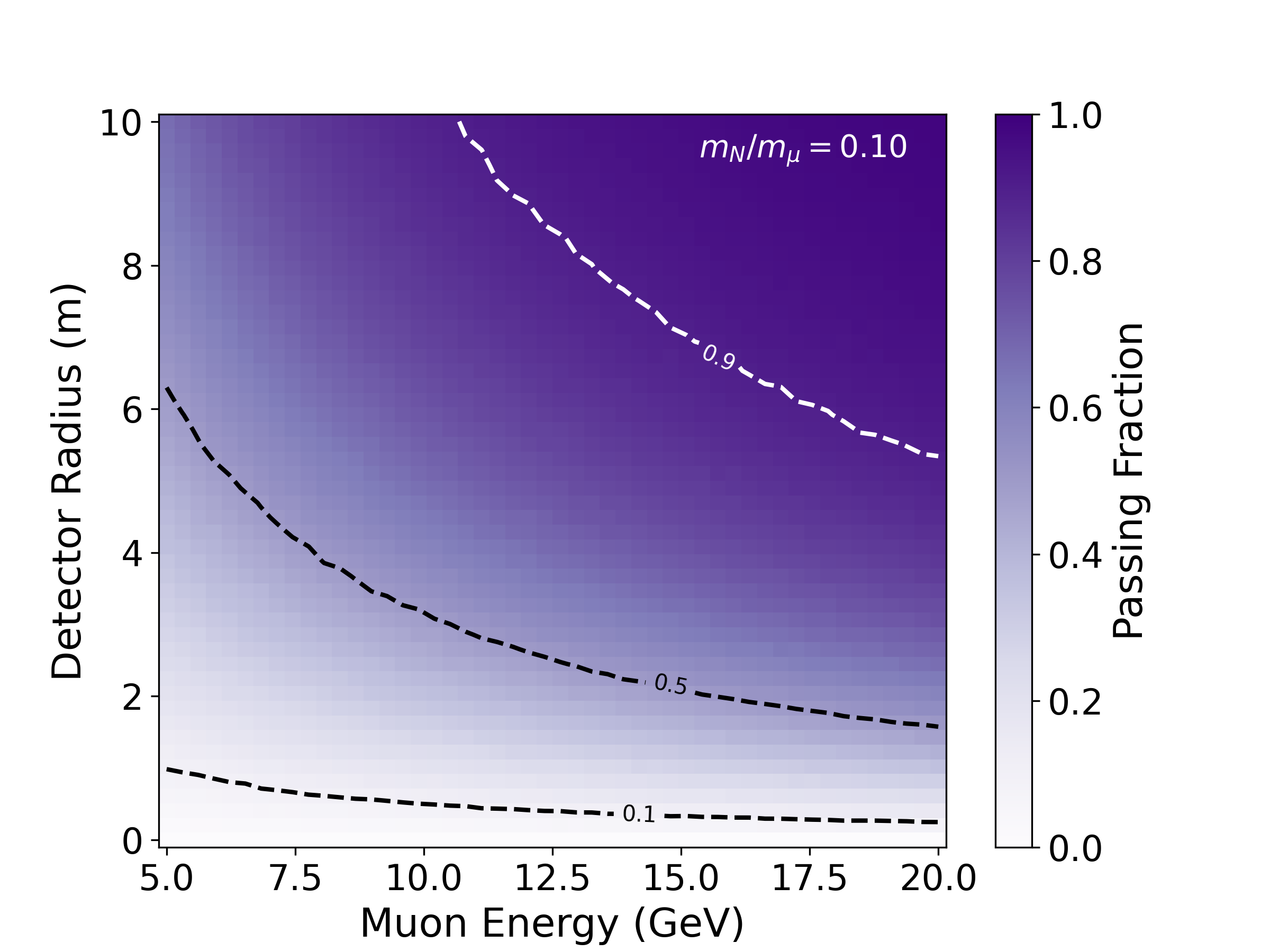}
    \caption{The dependence of $\epsilon_{\rm det}$ on the detector radius and muon energy, for muon-mixed HNLs with mass $m_\mu / 100$ (top) and $m_\mu / 10$ (bottom).}
    \label{fig:passing_fractions}
\end{figure}

The passing fraction $\epsilon_{\rm det}$ is computed by Monte Carlo: HNLs are sampled in energy and isotropic direction in the muon rest frame following \cref{eq:mu_diff_width}, boosted to the lab frame, and $\epsilon_{\rm det}$ is taken as the fraction whose transverse displacement at the detector, $L\sin\theta_{\rm lab}$, falls within the detector radius.
In \cref{fig:passing_fractions} we show $\epsilon_{\rm det}$ as a function of the detector radius and muon energy, for two different HNL masses.
The acceptance grows with both the detector radius and the muon energy, the latter because a larger boost collimates the HNLs around the beam axis.
For the remainder of this work, we assume a detector radius of $5~{\rm m}$.

\begin{figure}
    \centering
    \includegraphics[width=\linewidth]{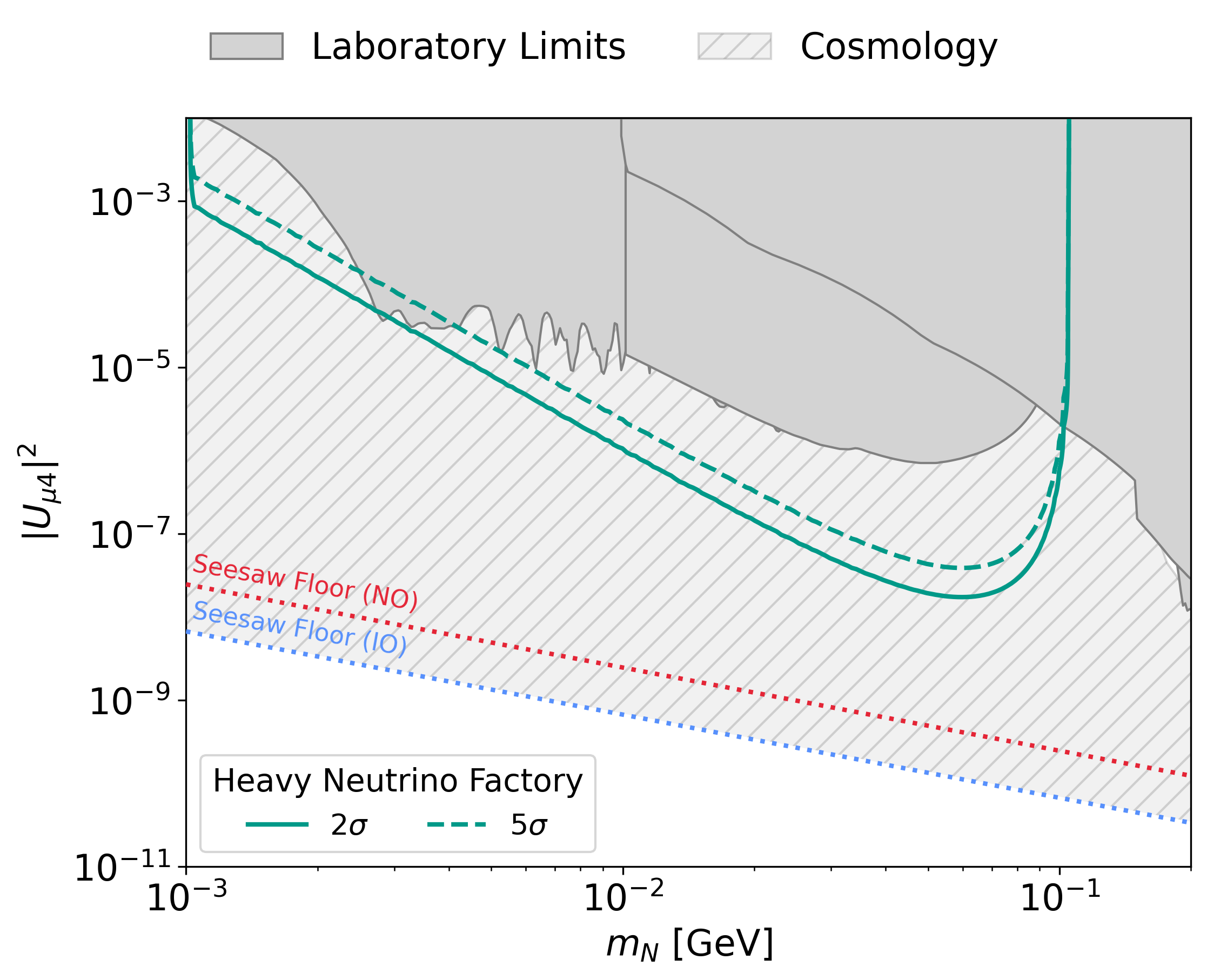}
    \caption{Sensitivity of the heavy neutrino factory setup to $\UNsq{\mu}$. Existing laboratory limits in gray come from Ref.~\cite{Fernandez-Martinez:2023phj} and references therein, and the gray hatched regions indicates BBN bounds on HNLs~\cite{Sabti:2020yrt}.
    Red and blue dotted lines represent the minimum mixing compatible with neutrino oscillation data in the seesaw framework for the normal and inverted ordering, respectively.
}
    \label{fig:muon_decay_sensitivity}
\end{figure}

The resulting HNL decay yields across the muon-mixing parameter space are shown in \cref{fig:muon_decay_sensitivity}.
Treating the search as background-free and requiring $N_{ee} \gtrsim 3.8$ ($15$) events for a $2\sigma$ ($5\sigma$) reach, this configuration would improve upon the LSND limits of Ref.~\cite{Ema:2023buz} by about two orders of magnitude in $\UNsq{\mu}$ across the $m_N \lesssim m_\mu$ mass range.
Similar sensitivity is possible for $\UNsq{e}$, as shown in \cref{sec:muon_decay_electron}.

\begin{figure}
    \centering
    \includegraphics[width=\linewidth]{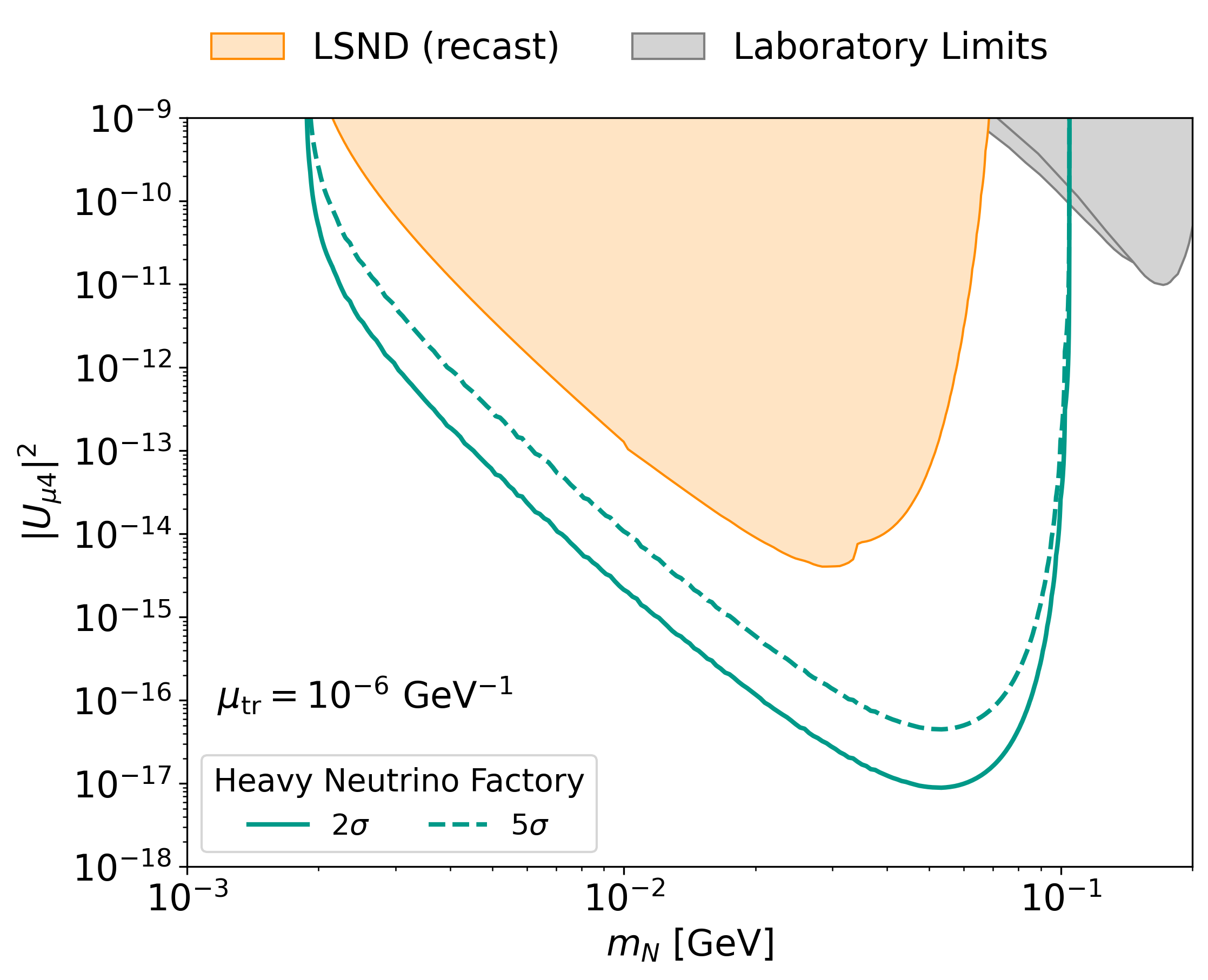}
    \caption{Sensitivity of the heavy neutrino factory to $\UNsq{\mu}$ under a model in which $N \to \nu e^+ e^-$ decays proceed via a transition magnetic moment $\mu_{\rm tr} = 10^{-6}~{\rm GeV}^{-1}$.}
    \label{fig:rates_dipole}
\end{figure}

As shown in \cref{fig:muon_decay_sensitivity}, the parameter space accessible in the heavy neutrino factory setup is already constrained by cosmology.
Under standard cosmological assumptions, HNLs with these mixings are populated in the early Universe and, if their lifetime $\tau_N\propto(G_F^2U^2m_N^5)^{-1}$ is long enough for them to survive into big-bang nucleosynthesis (BBN), their decays, mainly into active neutrinos, raise the neutron-to-proton ratio and hence the primordial $^4$He abundance~\cite{Dolgov:2000jw,Sabti:2020yrt}.
These bounds are avoided in extensions beyond the minimal mass mixing picture that shorten the HNL lifetime such that they decay before the onset of BBN.
As a representative example, we consider HNLs that couple to the active neutrinos through a transition magnetic moment $\mu_{\rm tr}$~\cite{Arguelles:2021dqn}.
A nonzero dipole coupling $\mu_{\rm tr}$ enhances the HNL decay rate, including the visible $N \to e^+ e^- \nu$ channel through an off-shell photon, thereby increasing the in-detector decay probability while production continues to proceed through mixing.
\Cref{fig:rates_dipole} shows the corresponding decay yields for a benchmark $\mu_{\rm tr} = 10^{-6}~{\rm GeV}^{-1}$.
The heavy neutrino factory has sensitivity to $\mu_{\rm tr}$ multiple orders of magnitude beyond existing limits, including a recast of LSND constraints from Ref.~\cite{Ema:2023buz}.

\section{Muon Beam Dump} \label{sec:beam_dump}

The muon beam-dump proposal uses muon scattering to produce HNLs that decay downstream of the target.
We study three configurations: a dense target with a low-density or evacuated decay volume, an Earth target with an atmospheric decay region, and an Earth target with an evacuated decay tunnel.

\subsection{Dense Target Plus Low-Density Decay Region} \label{sec:dump}

Analogously to the NA64$\mu$ experiment~\cite{Gertsenberger:2023ree}, a high-energy muon beam can be used to search for HNLs in a beam-dump configuration~\cite{Cesarotti:2022ttv}.
In this setup, the muon beam is dumped into a dense material such as tungsten, and the resulting large muon--nucleus luminosity compensates for the reduced center-of-mass energy, allowing the experiment to probe
mixing angles several orders of magnitude below the reach of collider searches across a wide range of HNL masses.

\subsubsection{HNL Production and Decay}
\begin{figure}
    \centering
    \includegraphics[width=\linewidth]{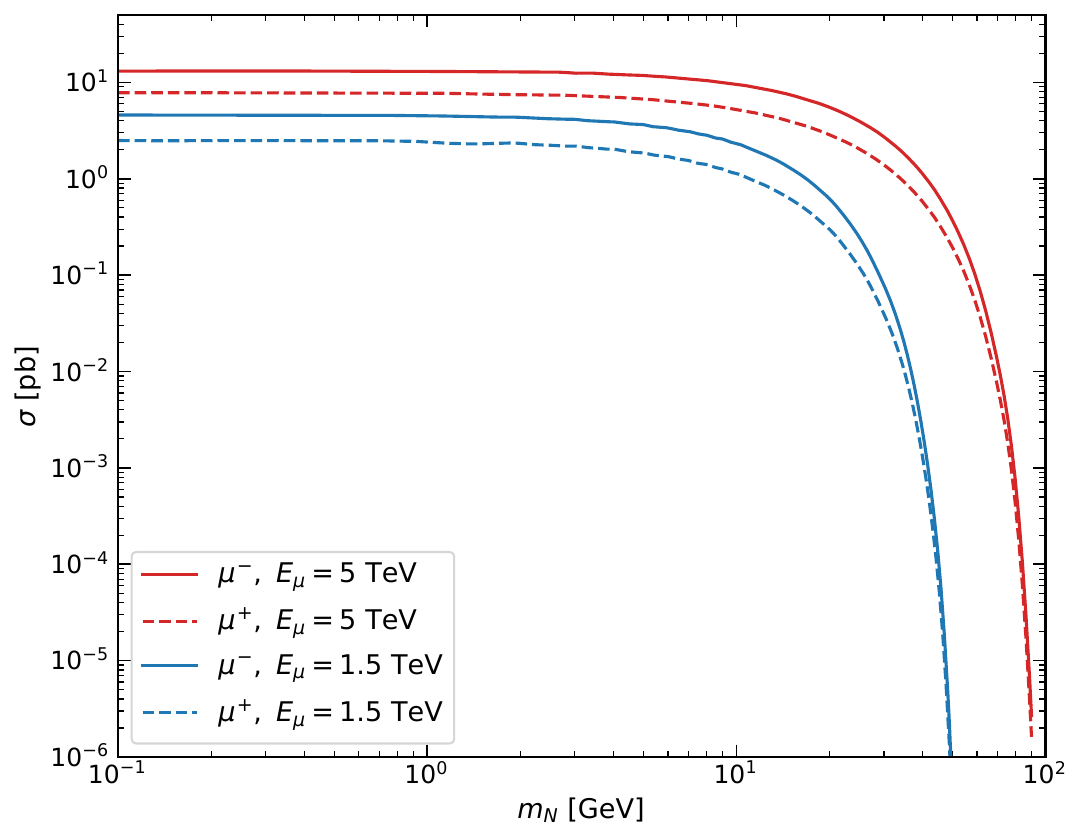}
    \caption{HNL production cross section $\sigma(\mu^\pm I \to N X)$ as a function of $m_N$ for $\mu^-$ and $\mu^+$ beams at $E_\mu = 1.5~\tev$ and $5~\tev$, normalized to $\UNsq{\mu} = 1$.
    Cross sections are computed with \texttt{MadGraph5\_aMC@NLO} for an isoscalar target based on the CT18NNLO PDFs~\cite{Hou:2019efy}.}
    \label{fig:XSMuDump}
\end{figure}

The primary production mechanism is deep-inelastic scattering, $\mu^{\pm} I \to N X$, through $t$-channel $W$-boson exchange, in which the HNL inherits the muon flavor through the mixing parameter $\UNsq{\mu}$.
In \cref{fig:XSMuDump}, we present the HNL production cross section as a function of the HNL mass $m_N$, with the mixing normalized to $\UNsq{\mu}=1$.
The incident muon energy is taken as $E_{\mu}=1.5~\tev$ and $5~\tev$, corresponding to the $3~\tev$ and $10~\tev$ muon collider configurations~\cite{InternationalMuonCollider:2024jyv}.
The calculation is performed with \texttt{MadGraph5\_aMC@NLO}~v2.9.13~\cite{Alwall:2014hca}, with the target modeled as an isoscalar using the CT18NNLO parton distribution functions~\cite{Hou:2019efy}.
The cross section is nearly independent of $m_N$ at low mass and falls steeply as $m_N$ approaches the kinematic limit $\sqrt{s} - m_p \approx \sqrt{2 m_p E_\mu} \approx 53~(97)~\gev$ at $E_\mu = 1.5~(5)~\tev$.
The $\mu^+$ cross sections are smaller than the $\mu^-$ ones by a factor of about $1.5$, as the antilepton beam scatters off valence quarks with a helicity suppression.

\begin{figure}
    \centering
    \includegraphics[width=\linewidth]{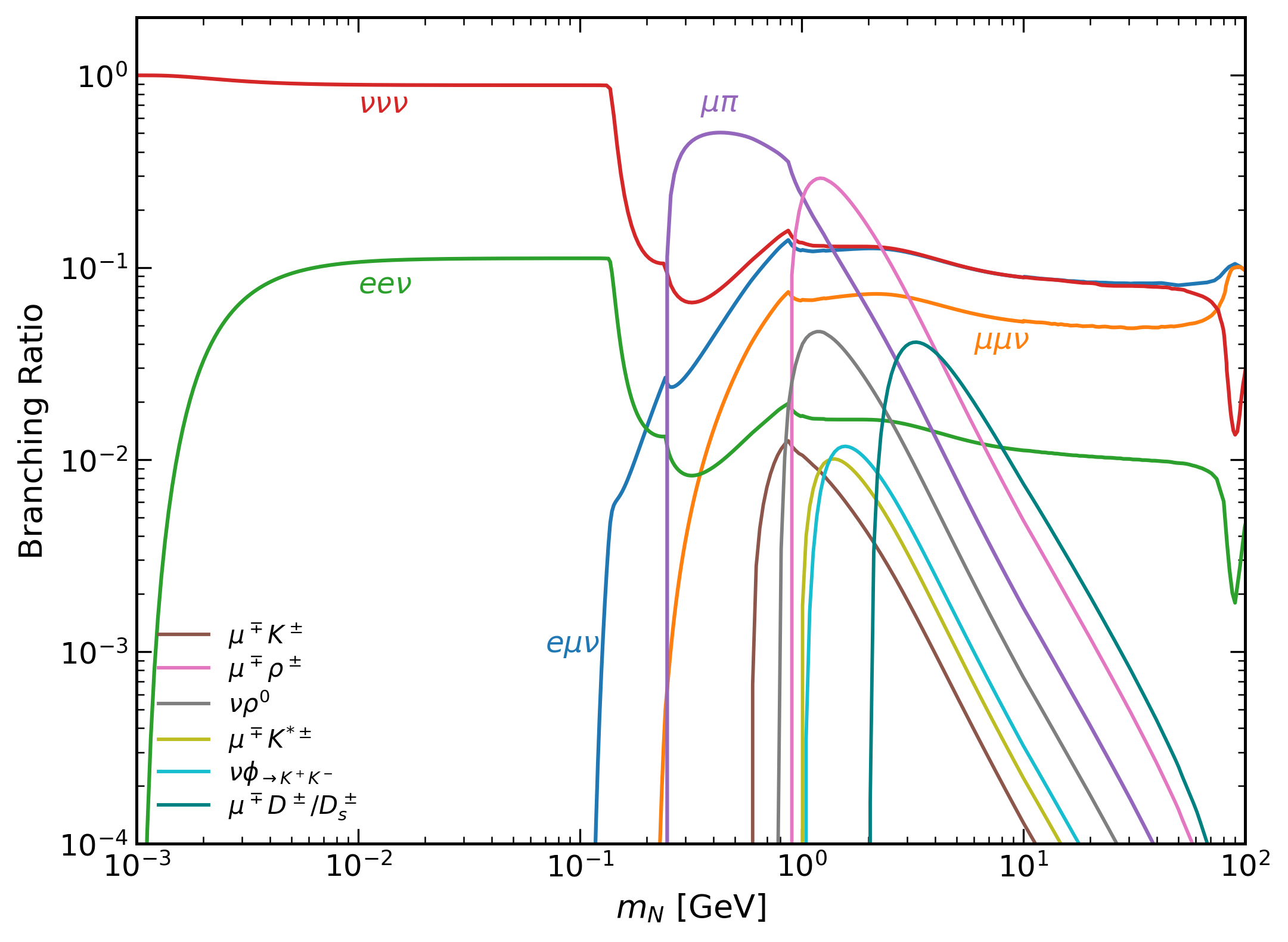}
    \caption{HNL decay branching fractions as a function of $m_N$ for pure muon mixing ($\UNsq{\mu} \neq 0$, $\UNsq{e} = \UNsq{\tau} = 0$): the invisible channel $N \to 3\nu$ and all exclusive channels with two charged tracks in the final state.}
    \label{fig:BRmuDump}
\end{figure}

HNLs produced in the target propagate downstream and decay inside the detector volume.
For the scenario of pure muon mixing ($\UNsq{\mu} \neq 0$, $\UNsq{e} = \UNsq{\tau} = 0$), \cref{fig:BRmuDump} shows the branching fractions of the invisible channel $N \to 3\nu$ and of every decay channel with exactly two charged tracks in the final state. At low mass, the invisible channel accounts for nearly the entire width, and the only visible channel is $N \to \nu e^+e^-$, which carries a branching fraction of about $10\%$ for $m_N \lesssim m_{\mu} \approx 0.11~\gev$. As $m_N$ increases, the remaining dilepton channels open in turn: the lepton-flavor-violating channel $N \to e^{\pm}\mu^{\mp}\nu$ becomes accessible near $m_N \approx m_{\mu}$, and the $N \to \nu\mu^+\mu^-$ channel opens at the dimuon threshold $m_N \approx 2m_{\mu} \approx 0.21~\gev$. Semileptonic two-body channels follow: $N \to \mu^{\mp}\pi^{\pm}$ opens at $m_N \approx m_{\mu} + m_{\pi} \approx 0.25~\gev$ and quickly dominates the visible width, reaching a branching fraction of about $50\%$ near $m_N \approx 0.5~\gev$, followed at their respective thresholds by $\mu^{\mp}K^{\pm}$, $\mu^{\mp}\rho^{\pm}$, $\nu\rho^0$, $\mu^{\mp}K^{*\pm}$, $\nu\phi(\to K^+K^-)$, and $\mu^{\mp}D^{\pm}/D_s^{\pm}$. Since the exclusive two-body widths scale as $m_N^3$ while the total width grows as $m_N^5$, these branching fractions fall off as $\sim 1/m_N^2$ and become negligible above $\sim 10~\gev$. In this high-mass region the dilepton modes dominate the two-track final states: $N \to e^{\pm}\mu^{\mp}\nu$ settles at roughly $10\%$, $N \to \nu\mu^+\mu^-$ at  about $7\%$, and $N \to \nu e^+e^-$ at the percent level, while the invisible fraction falls to around $10\%$. The remainder of the width goes to hadronic final states with higher charged-track multiplicity, which are not shown.

\subsubsection{Experimental Geometry and Backgrounds}

Similar to the experimental setup in~\cite{Cesarotti:2022ttv}, we consider a
beam-dump geometry consisting of a dense target of length
$\ell_{\mathrm{tar}}$, immediately followed by a shield of length
$\ell_{\mathrm{sh}} = 10~\mathrm{m}$ that combines a strong magnetic field
with passive shielding, dedicated to removing the residual muon beam, and any charged backgrounds produced in the target. Beyond the shield lies a low-density decay volume of length
$\ell_{\mathrm{dec}} = 10^{2}~\mathrm{m}$. For the target material we
consider two benchmarks: tungsten (W) of length
$\ell_{\mathrm{tar}} = 40~\mathrm{m}$, similar to but longer than that employed in the Fermilab
beam-dump experiment DONUT~\cite{DONUT:2000fbd}, and water (H$_2$O) of
length $\ell_{\mathrm{tar}} = 10~\mathrm{m}$, as adopted
in~\cite{Cesarotti:2022ttv}. The dense tungsten target maximizes the HNL
production rate per unit length, while the water target emits a
substantially smaller neutrino flux into the downstream volume.
The decay volume, in turn, must contain as little material as possible,
since neutrinos produced in the dump can scatter on any residual medium
and fake a decay vertex. Two designs have been established by past and
planned hidden-particle searches: the SHiP~\cite{SHiP:2021nfo,Miano:2021kyu}
and NA62~\cite{NA62:2023nhs,NA62:2023qyn} experiments employ evacuated
decay vessels, whereas the NuTeV decay channel was instrumented with
helium-filled bags at atmospheric
pressure~\cite{NuTeV:1999kej,NuTeV:2001ndo}. We consider both options:
a helium-filled volume at $1$~atm serves as our baseline, for which the
neutrino-induced background is estimated explicitly, and we then show
that evacuating the volume to a SHiP/NA62-like pressure renders the
searches background-free.
Throughout, we assume $N_{\mu} = 10^{22}$ muons on target, a magnitude
comparable to that envisioned at a neutrino factory~\cite{Denton:2024glz}.

The expected number of inclusive HNLs signal events is obtained by integrating the production along the target, in which the muon loses energy,
\begin{equation}\label{eq:Nsigdump}
\begin{split}
    S =  & N_{\mu}\,n_{\rm nucl}\,\UNsq{\mu}\,\epsilon_{\rm det}
           \int_0^{\ell_{\rm tar}} \! dx\; \sigma_0\big(E_\mu(x),\, m_N\big) \\
         & \times e^{-(\ell_{\rm tar} - x + \ell_{\rm sh})/\ell_d(x)}
           \left(1 - e^{-\ell_{\rm dec}/\ell_d(x)}\right),
\end{split}
\end{equation}
where $n_{\rm nucl} = \rho N_A$ is the nucleon number density of the target, with $\rho$ its mass density and $N_A$ Avogadro's number, $x$ is the depth in the target, $E_\mu(x)$ is the muon energy at that depth, $\sigma_0(E_\mu, m_N)$ is the HNL production cross section per nucleon, normalized to unit mixing, and $\ell_d(x) \propto 1/\UNsq{\mu}$ is the mean HNL decay length, evaluated at the mean HNL energy for the local muon energy. The exponential is the probability that the HNL survives the remainder of the target and the shield, and the last factor is the probability that it decays inside the decay volume. For a constant muon energy \cref{eq:Nsigdump} reduces to the expression of Ref.~\cite{Cesarotti:2022ttv}. The detector efficiency $\epsilon_{\rm det}$ is estimated by Monte Carlo simulation as the fraction of events emitted at an angle $\theta < \theta_{\rm max} \equiv R_{\rm det} / L_{\rm det}$, where $R_{\rm det}$ is the detector radius and $L_{\rm det}$ is the distance from the production region to the detector.

The muon energy follows from the continuous energy loss $-dE_\mu/dX = a + bE_\mu$, with $X = \rho x$ the column depth, which gives $E_\mu(x) = (E_0 + a/b)\,e^{-b\rho x} - a/b$ for a muon injected with energy $E_0$ (see also \cref{eq:dEdx,eq:muon_range}).
We use $a = 2.1\times10^{-3}~{\rm GeV/(g\,cm^{-2})}$ and $b = 1.6\times10^{-5}~{\rm (g\,cm^{-2})^{-1}}$ for tungsten and $a = 3.1\times10^{-3}~{\rm GeV/(g\,cm^{-2})}$ and $b = 3.1\times10^{-6}~{\rm (g\,cm^{-2})^{-1}}$ for water, the averages of the tabulated values over $E_\mu = 1$--$5~\tev$~\cite{Groom:2001kq}.
A muon of $5~(1.5)~\tev$ leaves the $40~{\rm m}$ tungsten target, a column depth of $7.7\times10^{4}~{\rm g\,cm^{-2}}$, with $1.4~(0.34)~\tev$, whereas in the water target it loses less than $1\%$ of its energy.
The cross section at the injection energy is that of \cref{fig:XSMuDump}, and its dependence on the muon energy is that of \cref{fig:energy_dep_production_xs}.

The corresponding background originates from two sources: Standard Model neutrinos from muon decays in flight within the target, and from muon--nucleus scattering in the target, which subsequently interact in the decay volume. In the following, we adopt a tungsten target for illustration unless stated otherwise.

\begin{figure}[h]
    \centering
    \includegraphics[width=\linewidth]{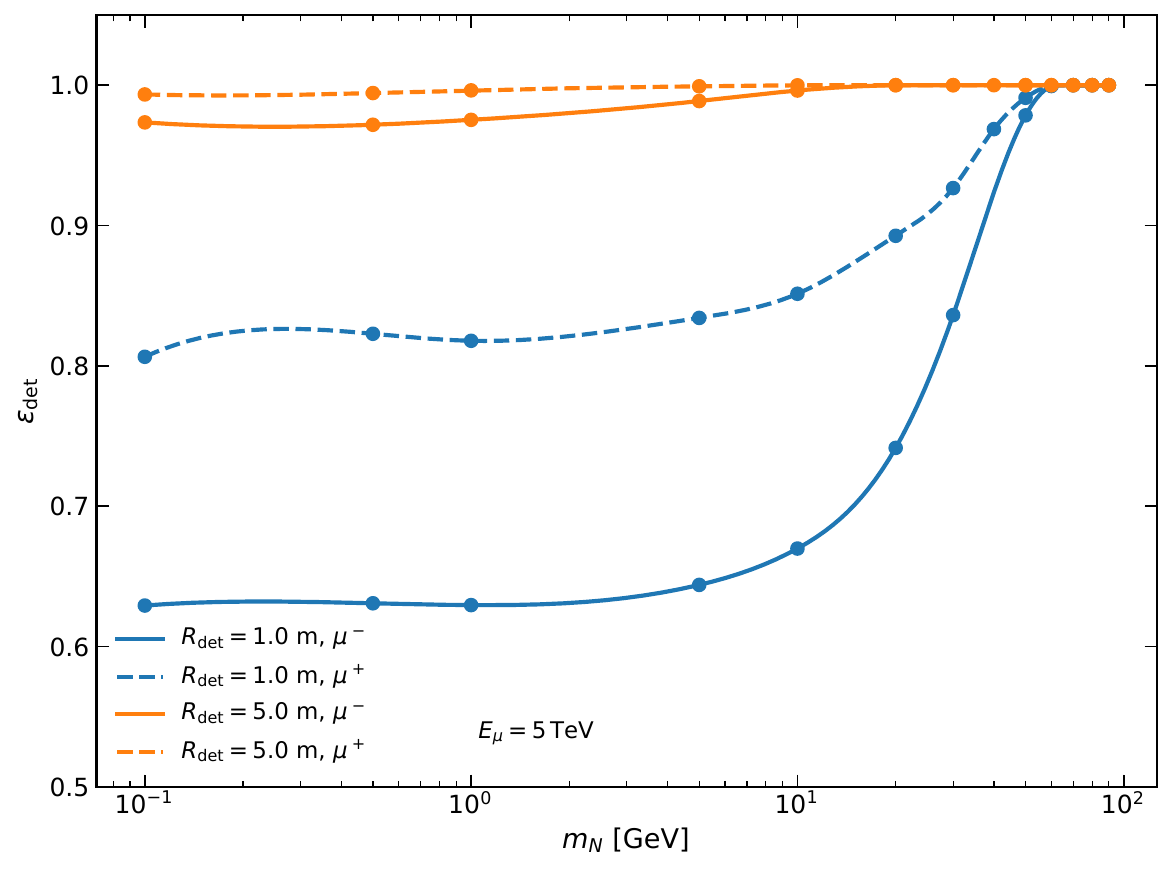}
    \caption{Detector efficiency as a function of HNL mass.}
    \label{fig:eff_mn}
\end{figure}

\begin{figure}
    \centering
    \includegraphics[width=\linewidth]{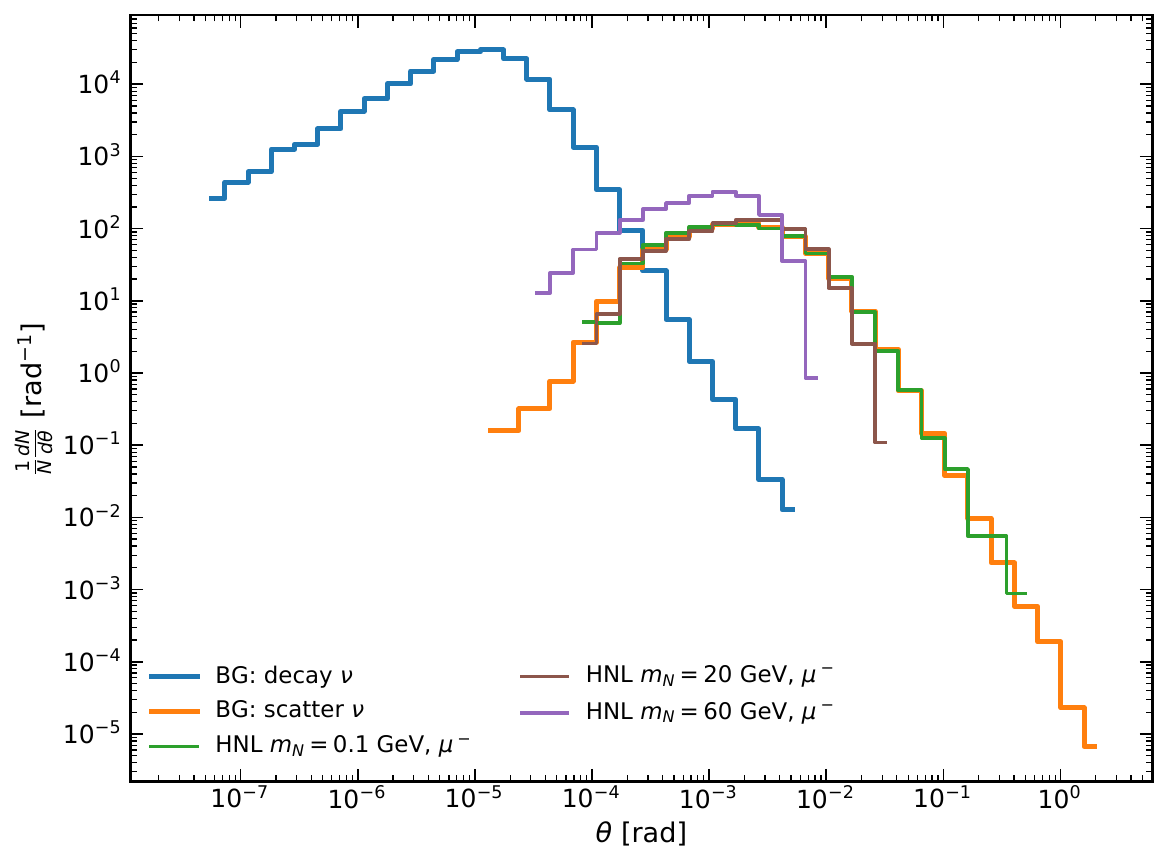}
    \includegraphics[width=\linewidth]{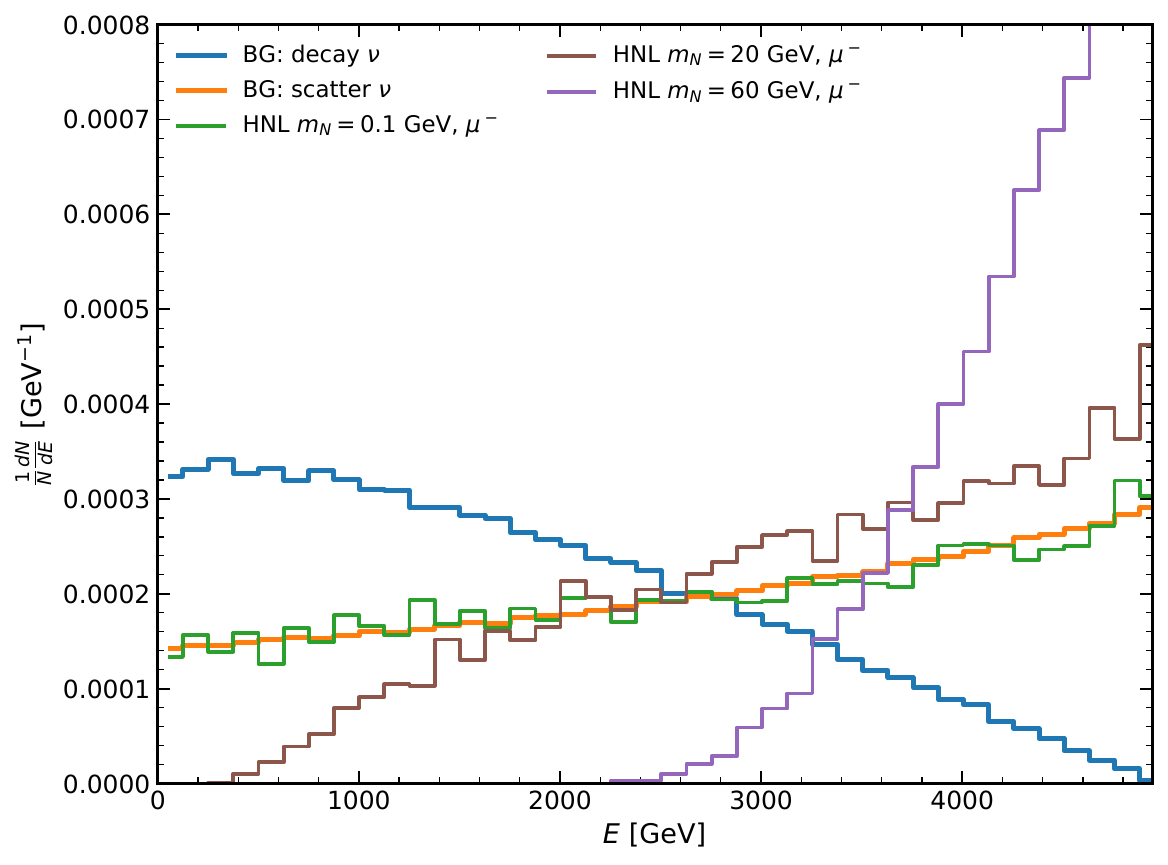}
    \caption{Normalized angular (upper) and energy (lower) distributions for background neutrinos from muon decays in flight (BG: decay $\nu$) and muon scattering (BG: scatter $\nu$), compared with HNL signals at benchmark masses of $m_N = 0.1$, $20$, and $60~\gev$ for a $\mu^-$ beam at $E_\mu = 5~\tev$.}
    \label{fig:E_theta_distribution}
\end{figure}

The angular and energy distributions of these two backgrounds, compared to HNL signal for several benchmark masses, are shown in \cref{fig:E_theta_distribution}.
In the angular distribution (upper panel), neutrinos from the decay background source are more forward, while neutrinos from the scattering background source have an angular distribution similar to the low-mass HNL signal, since both are produced via $t$-channel $W$-boson exchange.
The angle decreases as the HNL mass increases, because heavier HNLs carry a larger fraction of the muon beam momentum.
Representative values for several $R_{\rm det}$ are listed in \cref{tab:eff}. A detector radius of at least a few meters is required to maintain good signal acceptance. The dependence of the detector efficiency on $m_N$ is shown in \cref{fig:eff_mn}.

\begin{table}[h]
\centering
\caption{Detector efficiency $\epsilon_{\rm det}$ for several detector radii $R_{\rm det}$ and for signals and backgrounds at $E_{\mu^-} = 5~\tev$.}
\label{tab:eff}
\begin{tabular}{cccccc}
\hline\hline
$R_{\rm det}~[\mathrm{m}]$ & $\epsilon_{\rm det}(0.1)$ & $\epsilon_{\rm det}(20)$ & $\epsilon_{\rm det}(60)$ & $\epsilon_{\rm det}(\text{decay})$ & $\epsilon_{\rm det}(\text{scatter})$ \\
\hline
0.5 & 0.353 & 0.397 & 0.855 & 1.000 & 0.354 \\
1   & 0.629 & 0.742 & 0.999 & 1.000 & 0.630 \\
5   & 0.974 & 1.000 & 1.000 & 1.000 & 0.972 \\
\hline\hline
\end{tabular}
\end{table}

In the energy distribution (lower panel in \cref{fig:E_theta_distribution}), the neutrino flux from the decay background source decreases with increasing energy, while the scatter background and the HNL signal both peak at higher energies.

\begin{figure}
    \centering
    \includegraphics[width=\linewidth]{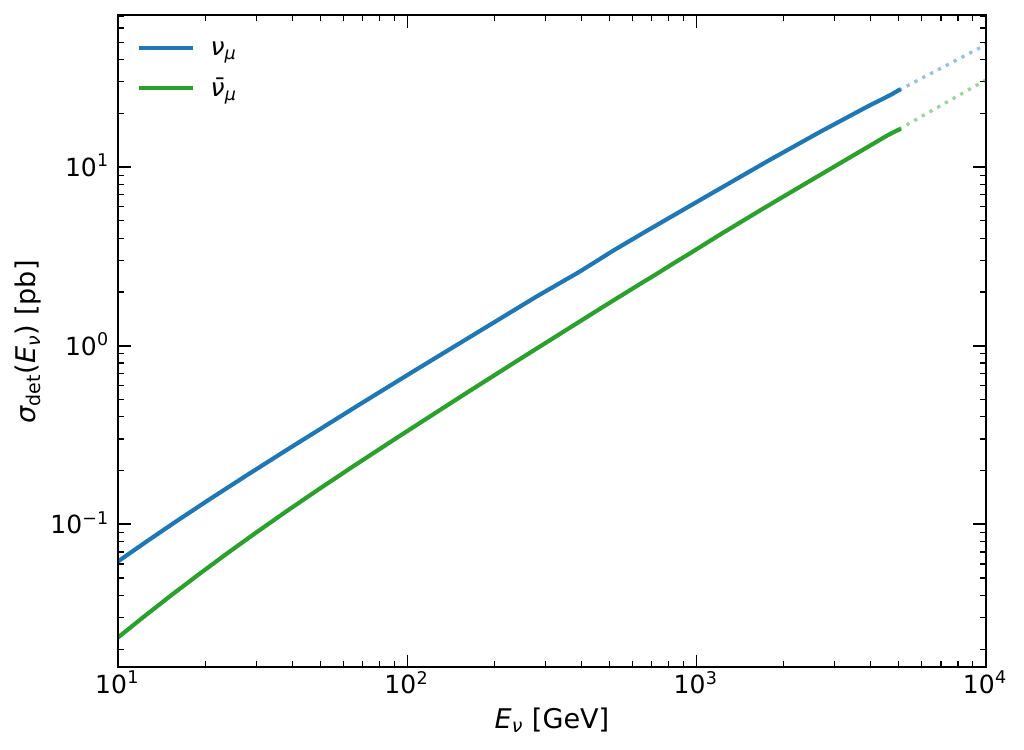}
    \caption{Total neutrino--nucleon interaction cross section $\sigma_{\nu \to \mu}(E_\nu)$ for background process in decay volume as a function of neutrino energy.}
    \label{fig:xs_E}
\end{figure}
The neutrino--nucleon interaction cross section $\sigma_{\nu \to \mu}(E_\nu)$ used in the background estimate is shown in \cref{fig:xs_E}.
\begin{table}[h]
\centering
\caption{Total detected background event counts for the W--He and H$_2$O--He target--detector configurations, for $\mu^-$ and $\mu^+$ beams at $E_\mu = 1.5~\tev$ and $5~\tev$, assuming $N_\mu = 10^{22}$ muons on target and $R_{\rm det} = 5~\mathrm{m}$. W--He geometry: $(\ell_{\rm tar},\,\ell_{\rm sh},\,\ell_{\rm dec}) = (40,\,10,\,100)~\mathrm{m}$. H$_2$O--He geometry: $(10,\,10,\,100)~\mathrm{m}$.}
\label{tab:bgevents}
\begin{tabular}{lcccc}
\hline\hline
Configuration & $5~\tev~\mu^-$ & $5~\tev~\mu^+$ & $1.5~\tev~\mu^-$ & $1.5~\tev~\mu^+$ \\
\hline
W--He \;[$\times 10^5$]      & 2.48 & 1.24 & 1.68 & 0.87 \\
H$_2$O--He \;[$\times 10^4$] & 3.72 & 2.09 & 3.93 & 2.09 \\
\hline\hline
\end{tabular}
\end{table}
The total background event counts for both target configurations and all four beam conditions are summarized in \cref{tab:bgevents}, assuming $R_{\rm det} = 5~\mathrm{m}$.
For the tungsten target with a helium decay volume, the $\mu^-$ beam yields $B \approx 2.5$ $\times 10^5$ ($1.7\times 10^5$) events at 5 (1.5) TeV, while the $\mu^+$ beam produces roughly half as many.
Replacing the tungsten target with water reduces the background by a factor of about $4-7$. The water-target $\mu^-$ ($\mu^+$) background is $B \approx 3.7 \times 10^4$ ($2.1 \times 10^4$) at $5~\tev$ and $B \approx 3.9 \times 10^4$ ($2.1 \times 10^4$) at $1.5~\tev$.

\subsubsection{Sensitivity}
We quantify the projected reach as the median expected exclusion obtained with the modified-frequentist CL$_s$ method~\cite{Read:2002hq,Segura:2024prac,ParticleDataGroup:2024cfk}. For a background-free search, the $2\sigma$ and $5\sigma$ sensitivities correspond to signal yields of $S_{2\sigma} \simeq 3.8$ and $S_{5\sigma} \simeq 15$ events, whereas for searches with sizable background we estimate the significance as $S/\sqrt{S + B}$.

\begin{figure}[!h]
    \centering
    \includegraphics[width=\linewidth]{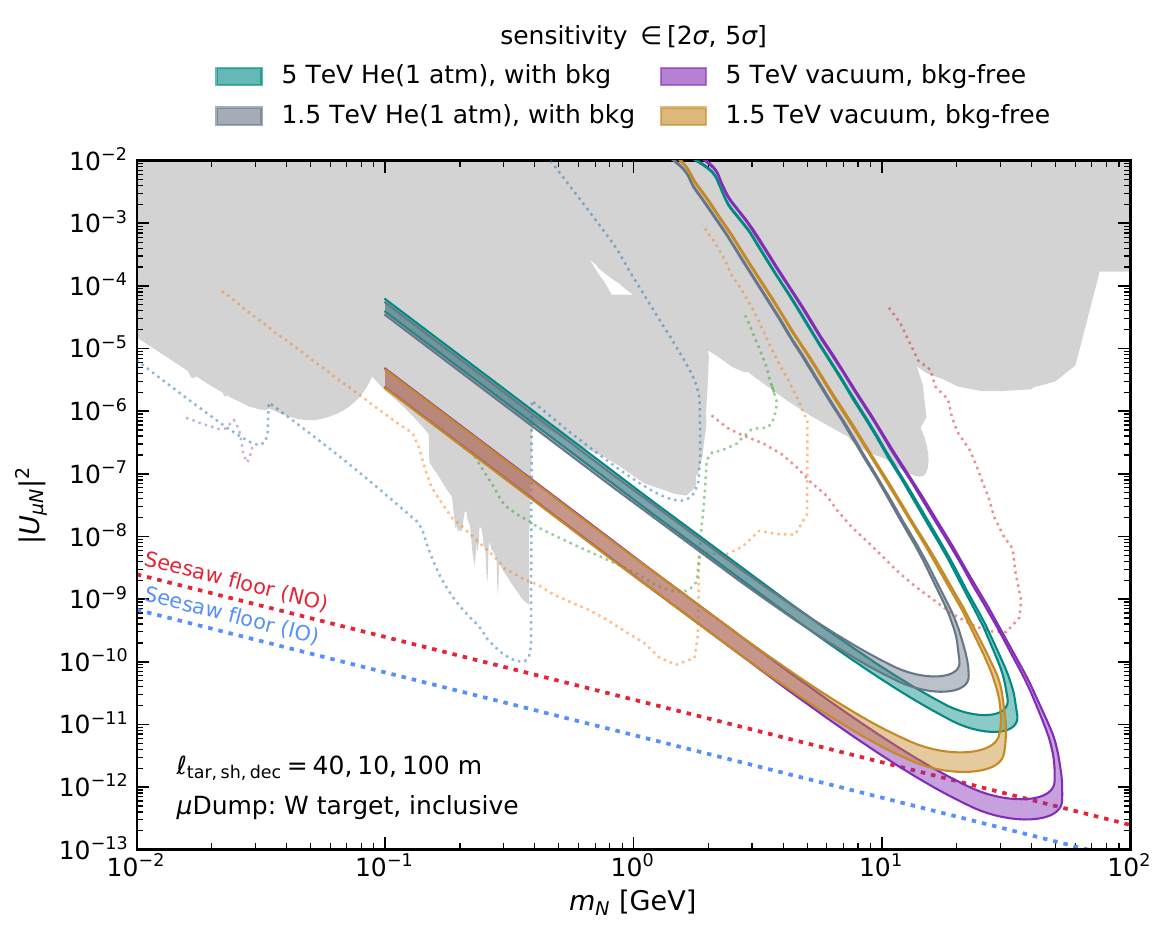}
    \caption{Projected sensitivity in the $(m_N, \UNsq{\mu})$ plane from the muon-dump experiment using the inclusive signal for a $\mu^-$ beam.
    Each shaded band spans the sensitivity range between the $2\sigma$ (outer) and $5\sigma$ (inner) contours.
    Teal ($5~\tev$) and slate-gray ($1.5~\tev$) bands correspond to the helium-filled decay volume and include the backgrounds of \cref{tab:bgevents} via $S/\sqrt{S+B}$; purple ($5~\tev$) and amber ($1.5~\tev$) bands correspond to the evacuated, background-free decay volume, \cref{eq:vacP}.
    Gray filled regions are current experimental limits~\cite{WA66:1985mfx,NA3:1986ahv,CHARMII:1994jjr,Daum:1987bg,DELPHI:1996qcc,NuTeV:1999kej,CMS:2016aro,CMS:2018iaf,CMS:2018jxx,PIENU:2019usb,T2K:2019jwa,Arguelles:2021dqn,ATLAS:2019kpx,ATLAS:2025lva,Bryman:2019bjg,NA62:2021bji,Kelly:2021xbv,CMS:2022fut,ATLAS:2022atq}.
    Dotted colored curves indicate projected sensitivities from the DUNE near detector~\cite{Breitbach:2021gvv}, SHADOWS~\cite{NA62:2020upd}, SHiP~\cite{SHiP:2018xqw}, PIONEER~\cite{PIONEER:2022yag}, and HL-LHC~\cite{Drewes:2019fou}.}
    \label{fig:WsensDumpAll}
\end{figure}

\paragraph*{Tungsten target.}
The projected sensitivity in the $(m_N,\, \UNsq{\mu})$ plane for the tungsten configuration is shown in \cref{fig:WsensDumpAll}, using the inclusive signal in which every HNL decay is counted.
For the helium-filled decay volume, the teal ($5~\tev$) and slate-gray ($1.5~\tev$) bands carry the full background of \cref{tab:bgevents} through $S/\sqrt{S+B}$. Here, each shaded band spans the sensitivity range between its $2\sigma$ (outer) and $5\sigma$ (inner) contours, and only the $\mu^-$ beam is shown.
At $E_{\mu} = 1.5~\tev$, the $2\sigma$ reach extends to $\UNsq{\mu}$ as small as $\sim 3.1\times10^{-11}$ for $m_N \sim 18~\gev$, improving to $\sim 6.9\times10^{-12}$ at $m_N \sim 29~\gev$ for $5~\tev$.
Both configurations already improve substantially upon the current experimental limits, shown as gray-filled regions~\cite{WA66:1985mfx,NA3:1986ahv,CHARMII:1994jjr,Daum:1987bg,DELPHI:1996qcc,NuTeV:1999kej,CMS:2016aro,CMS:2018iaf,CMS:2018jxx,PIENU:2019usb,T2K:2019jwa,Arguelles:2021dqn,ATLAS:2019kpx,ATLAS:2025lva,Bryman:2019bjg,NA62:2021bji,Kelly:2021xbv,CMS:2022fut,ATLAS:2022atq}.

This reach is, however, limited almost entirely by the neutrino-induced background, which scales linearly with the density of the medium in the decay volume and can therefore be eliminated by evacuating the vessel.
That such a search is genuinely background-free is established by the two experiments introduced above.
The SHiP decay vessel is specified to stay below $10^{-3}$~bar precisely to suppress neutrino interactions in the fiducial volume, and full simulation --- including impact-parameter pointing to the target and a hermetic surrounding background tagger --- yields fewer than $0.1$ expected background events for $2\times10^{20}$ protons on target~\cite{SHiP:2021nfo,Miano:2021kyu}.
NA62 operated in beam-dump mode confirms in data: its searches for hidden-particle decays into fully reconstructed two-track final states in the evacuated tank observed zero background events~\cite{NA62:2023nhs,NA62:2023qyn}.
Since the neutrino flux traversing our decay volume is two to three orders of magnitude below those of these proton beam dumps, adopting the same strategy here is conservative.
Scaling the backgrounds of \cref{tab:bgevents} linearly with pressure, the SHiP-style design criterion $B < 0.1$ over the full run requires
\begin{equation}\label{eq:vacP}
    P \;\lesssim\; 4.1\times10^{-4}~\left(2.6\times10^{-3}\right)~\mathrm{mbar}
\end{equation}
for the tungsten (water) configuration, intermediate between the SHiP specification ($\lesssim 1~\mathrm{mbar}$) and the routine NA62 operating vacuum ($\sim 10^{-6}~\mathrm{mbar}$), and thus well within demonstrated technology.

The resulting background-free inclusive sensitivity is shown by the purple ($5~\tev$) and amber ($1.5~\tev$) bands in \cref{fig:WsensDumpAll}.
The $2\sigma$ reach improves to $\UNsq{\mu} \sim 1.6\times10^{-12}$ at $m_N \sim 22~\gev$ for $1.5~\tev$ and $\sim 2.8\times10^{-13}$ at $m_N \sim 40~\gev$ for $5~\tev$ --- a factor of $19$ ($24$) beyond the corresponding helium-filled contours. At $5~\tev$ the inclusive contour lies below the muon-flavor seesaw floor of \cref{eq:flavor_floor} for normal ordering over $m_N \approx 12$--$52~\gev$, by up to a factor of $2.5$, and remains a factor of $1.5$ above the floor for inverted ordering.

\begin{figure}
    \centering
    \includegraphics[width=\linewidth]{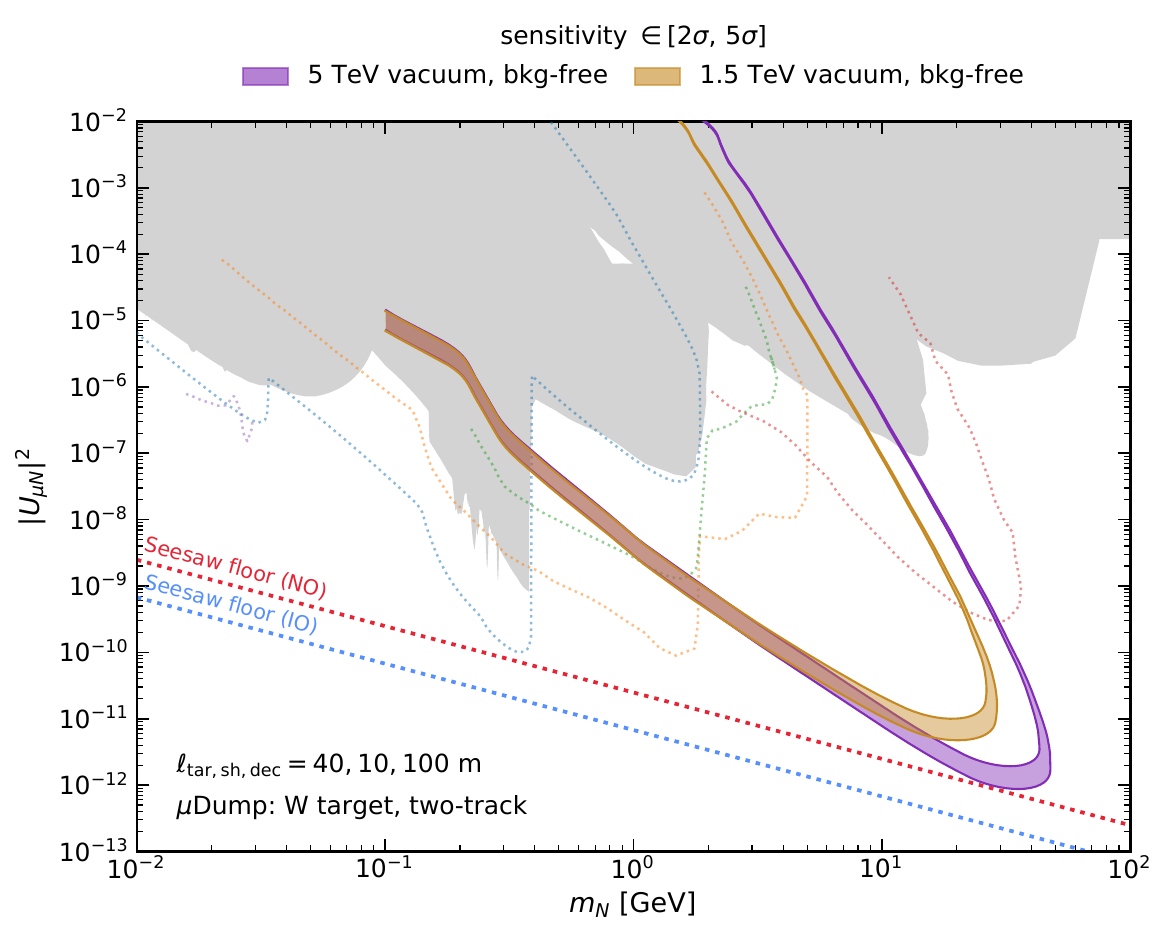}
    \caption{Same as \cref{fig:WsensDumpAll} but restricted to the exactly-two-charged-track decay channels of \cref{fig:BRmuDump}, in the evacuated (background-free) decay volume. The kink at $m_N \approx 0.25~\gev$ marks the $\mu\pi$ threshold.}
    \label{fig:WsensDumpDilep}
\end{figure}

The vacuum environment further allows us to adopt the same signal definition as SHiP and NA62: exactly two charged tracks forming a vertex in the fiducial volume and pointing back to the target.
This selection collects precisely the channels whose branching fractions are shown in \cref{fig:BRmuDump} --- the dilepton modes $\nu ee$, $\nu\mu\mu$, and $e\mu\nu$ together with the semileptonic two-body modes $\mu\pi$, $\mu K$, $\mu\rho$, $\nu\rho^0$, $\mu K^*$, $\nu\phi$, and $\mu D/D_s$ --- and trades some rate for a cleaner topology with full or partial mass reconstruction. The corresponding sensitivity is shown in \cref{fig:WsensDumpDilep}.
The $2\sigma$ reach is $\UNsq{\mu} \sim 4.4\times10^{-12}$ at $m_N \sim 20~\gev$ for $1.5~\tev$ and $\sim 8.0\times10^{-13}$ at $m_N \sim 36~\gev$ for $5~\tev$. 
At high masses the two-track reach is carried by the dilepton modes, while below $m_N \sim 2~\gev$ the $\mu\pi$ channel dominates the two-track rate, visible as the kink at its threshold $m_N \approx 0.25~\gev$.

\begin{figure}
    \centering
    \includegraphics[width=\linewidth]{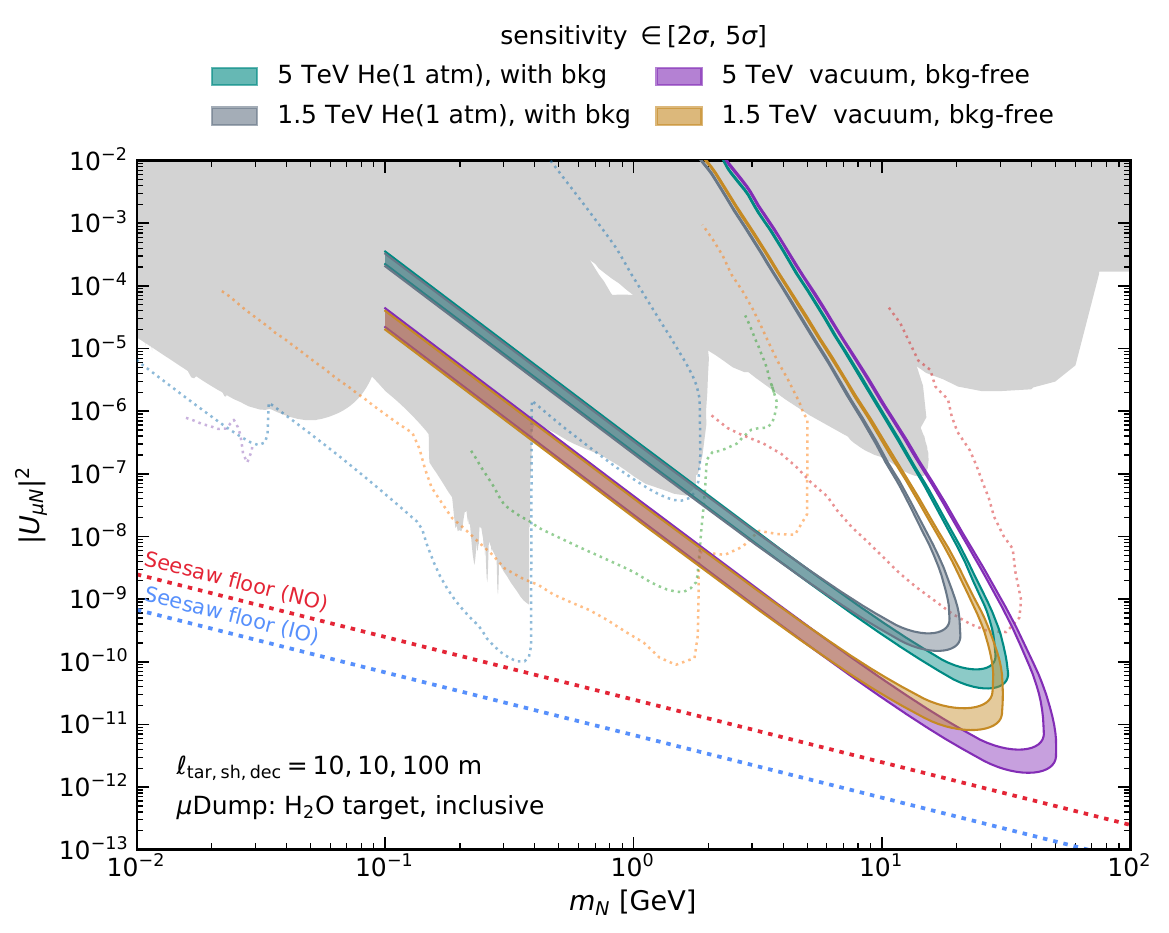}
    \caption{Same as \cref{fig:WsensDumpAll} but for a water target with $\ell_{\rm tar} = 10~{\rm m}$.}
    \label{fig:H2OsensDumpAll}
\end{figure}

\begin{figure}
    \centering
    \includegraphics[width=\linewidth]{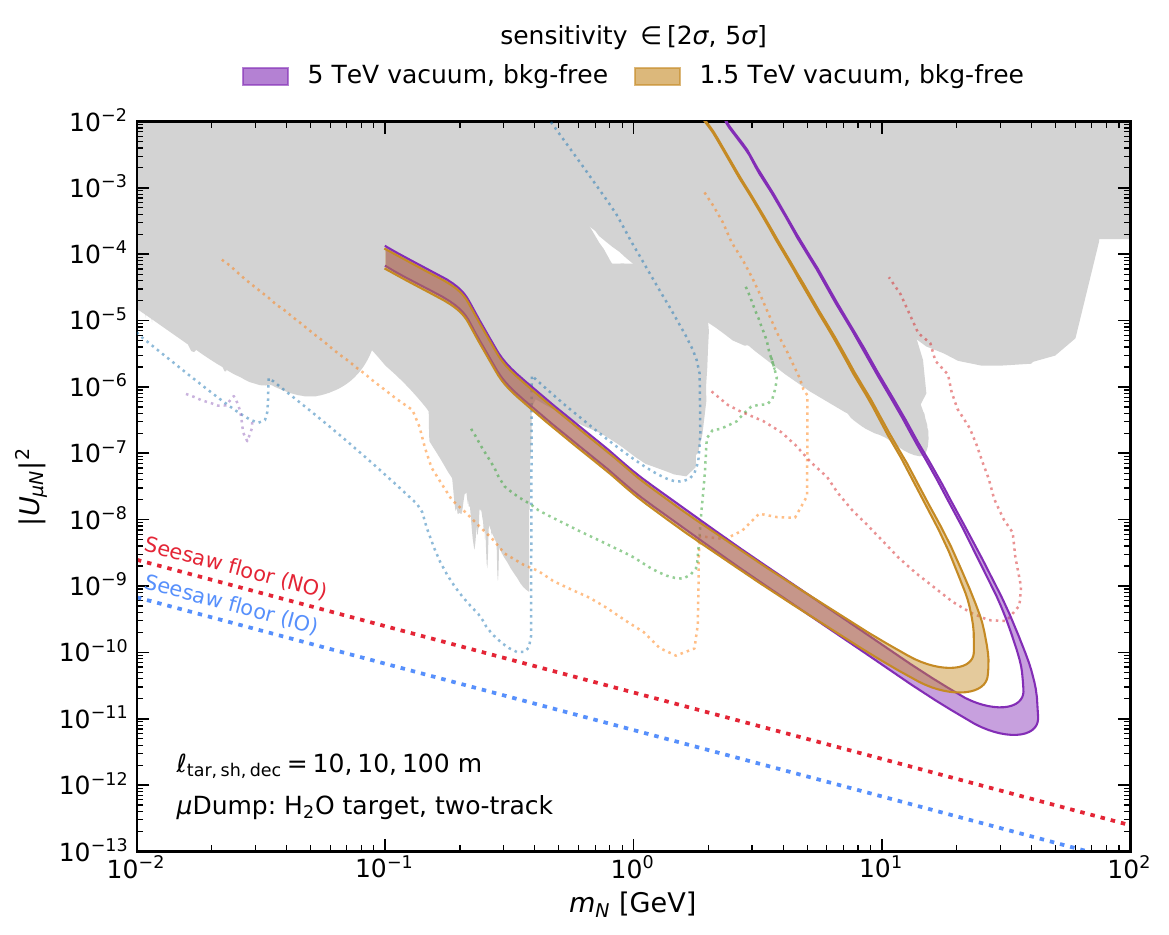}
    \caption{Same as \cref{fig:WsensDumpDilep} (two-track channels, evacuated decay volume) but for a water target with $\ell_{\rm tar} = 10~{\rm m}$.}
    \label{fig:H2OsensDumpDilep}
\end{figure}

\paragraph*{Water target.}
We also consider replacing the tungsten target with a water target of length $\ell_{\mathrm{tar}} = 10~\mathrm{m}$, which reduces the beam-induced neutrino background by a factor of four to seven (\cref{tab:bgevents}) and correspondingly relaxes the vacuum requirement of \cref{eq:vacP} to $2.6\times10^{-3}~\mathrm{mbar}$.
The projected sensitivities for the inclusive and two-track searches are shown in \cref{fig:H2OsensDumpAll,fig:H2OsensDumpDilep}, respectively, with the same conventions as for tungsten.
With the helium-filled decay volume, the inclusive $2\sigma$ reach of the $\mu^-$ beam extends to $\UNsq{\mu} \sim 1.4\times10^{-10}$ at $m_N \sim 17~\gev$ for $E_{\mu} = 1.5~\tev$, improving to $\sim 3.4\times10^{-11}$ at $m_N \sim 27~\gev$ for $5~\tev$, and evacuating the volume pushes these to $\sim 7.5\times10^{-12}$ at $m_N \sim 25~\gev$ and $\sim 1.5\times10^{-12}$ at $m_N \sim 40~\gev$, a gain of a factor of $18$ ($22$).
The background-free two-track search reaches $\UNsq{\mu} \sim 2.3\times10^{-11}$ at $m_N \sim 22~\gev$ for $1.5~\tev$ and $\sim 5.1\times10^{-12}$ at $m_N \sim 36~\gev$ for $5~\tev$.
The water-target reach remains weaker than the tungsten case at high $m_N$, where the production rate in the shorter, less dense target is the limiting factor. Its advantages are the smaller neutrino flux and the less demanding vacuum specification.

\subsection{Earth Target Plus Atmospheric Decay Region} \label{sec:balloon}

To maximize the reach of a muon beam dump, we consider a configuration in which the beam dump target is the Earth and the decay volume is the atmosphere.
In this setup HNLs are detected via their decays to pairs of muons that emit Cherenkov light in air, where $(n_{\rm air} \approx 1.0003)$.
To produce HNLs, the muon beam is steered into the Earth at a near-horizontal angle, traversing approximately two kilometers of bedrock and absorbing all muons before re-emerging into the atmosphere.
The bedrock serves as a dense, high-luminosity target in which muons upscatter into HNLs, while the low-density atmosphere serves as an enormous decay region in which the HNLs decay to a visible final state.
The Cherenkov light from the charged decay products is imaged by an air-Cherenkov camera of the type developed for EUSO-SPB2~\cite{Eser:2023lck}, POEMMA~\cite{POEMMA:2020ykm}, and Trinity~\cite{Otte:2025dld}.
Such a camera can either be suspended by a tethered balloon aerostat at an altitude of $h \sim 5~{\rm km}$, similar to the proposal in Ref.~\cite{Scrandis:2024bnb}, or situated atop a mountain as in Trinity~\cite{Otte:2025dld}.
Crucially, the near-horizontal geometry exploits the curvature of the Earth: a beam launched tangentially reaches the detector altitude only after traveling a distance $d \approx \sqrt{2 R_\oplus h} \approx 250~{\rm km}$, providing a decay baseline up to two orders of magnitude longer than the detector altitude itself.

This atmospheric muon beam dump setup must contend with neutrino backgrounds muon decays and scattering in the Earth, analogous to those in \cref{fig:E_theta_distribution}.
Despite the low density of the atmosphere, this background rate can be substantial due to the large neutrino flux and long atmospheric path lengths.
A free-flying or satellite-borne platform at higher altitude would further suppress the residual neutrino-induced background due to the exponential decrease in atmospheric density.
However, one would have to consider the reduced fraction of time during which the camera is along the beamline.
Such a study lies beyond the scope of this work; we focus throughout on the mountain-based or tethered-aerostat camera.
We use a custom Monte Carlo pipeline to simulate the rate of observed HNL decays in atmospheric muon beam dump sensitivity setup, the details of which are given below.

\subsubsection{HNL Production in the Earth} \label{sec:balloon_hnl_production}

\begin{figure}
    \centering
    \includegraphics[width=\linewidth]{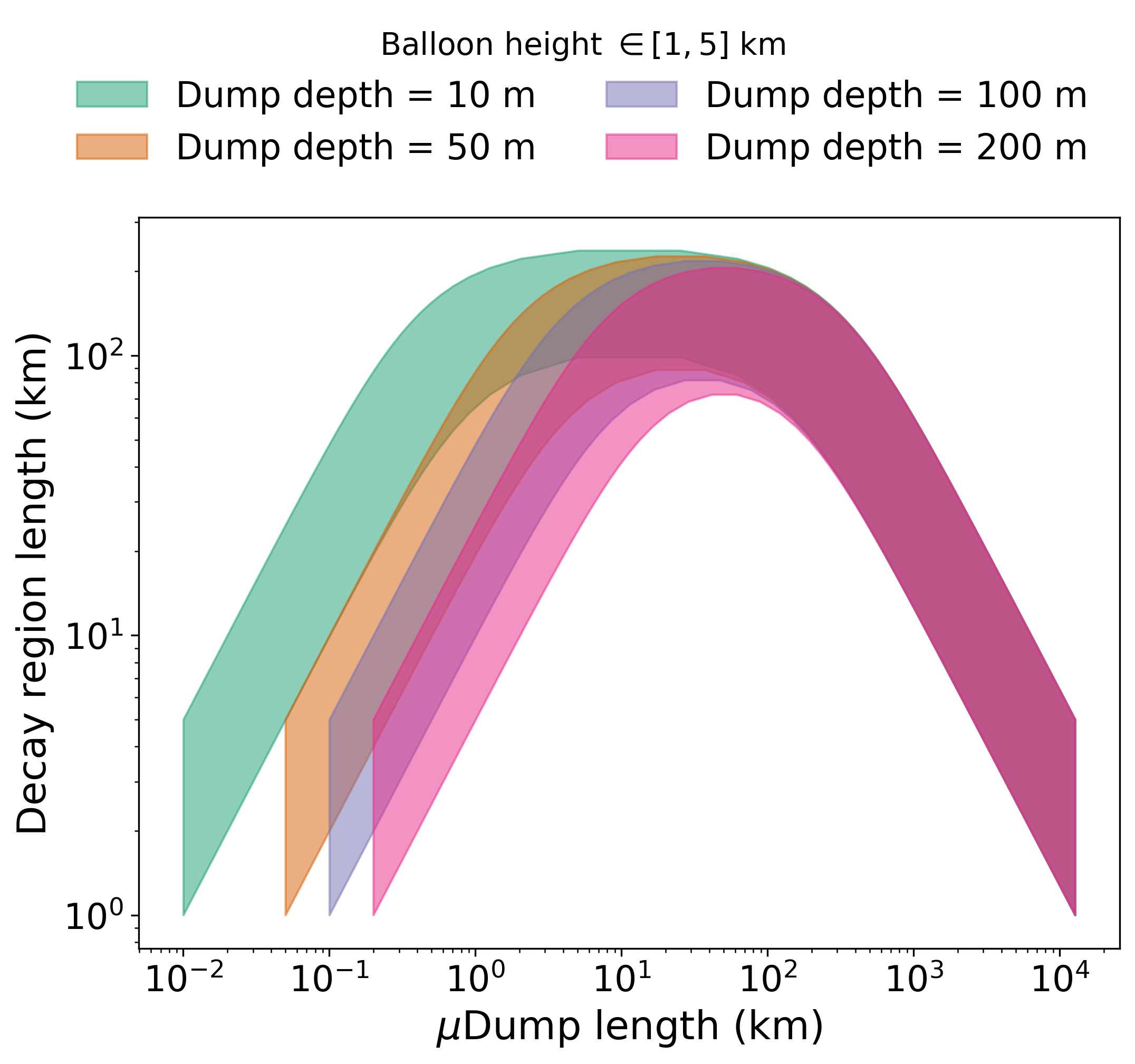}
    \caption{The possible atmospheric beam dump geometric configurations for a camera altitude between $1~{\rm km}$ and $5~{\rm km}$.
    Four different depths of the muon beam dump are shown.
    The envelopes for each are determined by sweeping over the angle of the muon beam with respect to the Earth's surface, from vertically upward (left edge) to vertically downward (right edge).}
    \label{fig:atmo_dump_geometry}
\end{figure}

In the atmospheric beam dump configuration, the Earth itself serves as a target for HNL production.
The muon beam is oriented along a nearly-horizontal direction with respect to the Earth's surface, such that the muon beam traverses a few km of bedrock before exiting the surface of the Earth.
The exact angle is determined by the depth of the muon collider, but for reasonable depths between 10\,m and 200\,m, one can achieve atmospheric decay regions of $\mathcal{O}({\rm 100\,km})$ with few-km muon dump lengths through Earth, as shown in \cref{fig:atmo_dump_geometry}.
As a baseline, we consider a muon beam dump at $100~{\rm m}$ depth oriented at an angle of $2.3^\circ$ above the horizon, corresponding to a $2.4~{\rm km}$ ``target'' region of bedrock traversed by the muons.

As the muons traverse the bedrock they lose energy according to
\begin{equation}\label{eq:dEdx}
-\frac{dE_\mu}{dX} = a + bE_\mu,
\end{equation}
where $X$ is the column depth traversed, $a = 2.0\times10^{-3}~{\rm GeV/(g\,cm^{-2})}$ parametrizes ionization losses, and $b = 3.5\times10^{-6}~{\rm (g\,cm^{-2})^{-1}}$ the radiative losses from bremsstrahlung and pair production.
We note that $a$ and $b$ vary softly with muon energy, and the values we have taken are characteristic for muons at or slightly below TeV energies~\cite{Groom:2001kq}.
\Cref{eq:dEdx} integrates to
\begin{equation} \label{eq:muon_range}
E_\mu(X) = (E_0 + E_\mu^c)\,e^{-bX} - E_\mu^c,
\end{equation}
with the ionization and radiative terms balancing at the critical energy $E_\mu^c = a/b \approx 571~{\rm GeV}$; for the multi-TeV beams considered here, radiative losses dominate over the $2.4~{\rm km}$ rock path.

\begin{figure}
    \centering
    \includegraphics[width=\linewidth]{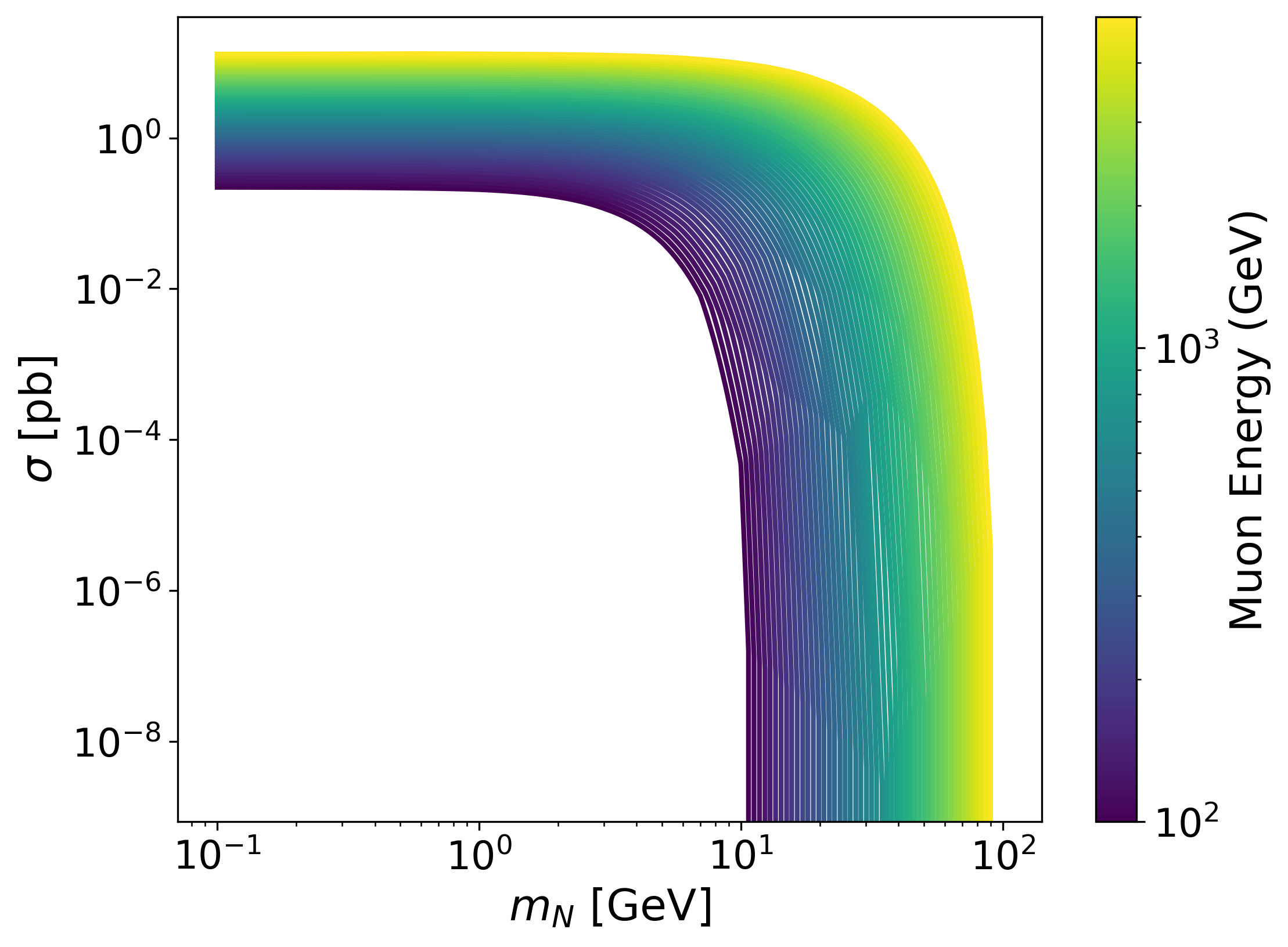}
    \caption{The $\mu I \to N X$ cross section for HNL production as a function of HNL mass, over a muon energies from $100~{\rm GeV}$ to $5~{\rm TeV}$.}
    \label{fig:energy_dep_production_xs}
\end{figure}

\begin{figure}
    \centering
    \includegraphics[width=\linewidth]{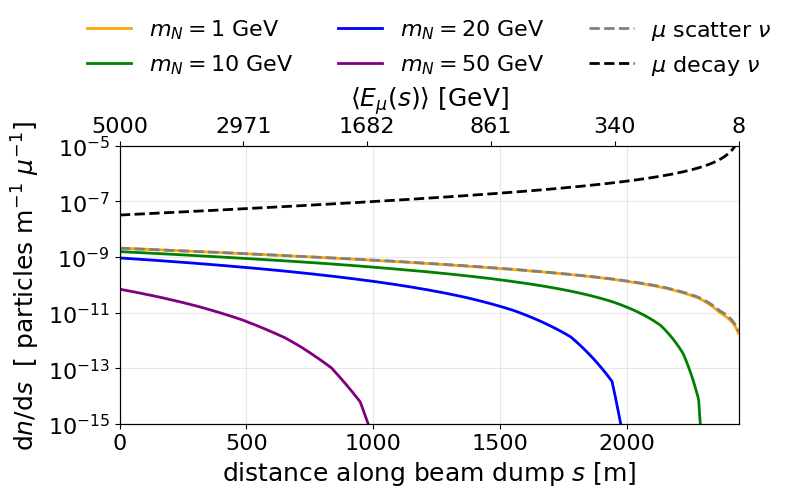}
    \caption{The production rate per muon as a function of distance $s$ along the Earth beamline, for four different HNL masses and both neutrino background sources.
    The top $x$ axis displays the average muon energy, which degrades as the muon traverses the Earth.
    }
    \label{fig:earth_production_pos}
\end{figure}

HNLs are produced along the muon track by deep-inelastic upscattering $\mu\, I \to N\, X$ through $t$-channel $W$ exchange.
As in \cref{sec:dump}, the cross section for this process is computed using \texttt{MadGraph5\_aMC@NLO} for an isoscalar target based on the CT18NNLO PDFs~\cite{Hou:2019efy}.
The caveat here is that we must integrate this cross section over the muon's degrading energy along the path.
\Cref{fig:energy_dep_production_xs} shows the HNL production cross section as a function of $m_N$ for a range of muon energies.
The number of HNLs produced by muon interactions in the Earth is then given by
\begin{equation} \label{eq:terrestrial_HNL_production}
    N_N = N_\mu \int_0^{\ell_{\rm max}} \sigma_{\mu I \to N X}(E_\mu(\ell)) n_{\oplus}d \ell
\end{equation}
where $N_\mu =10^{22}$ is the number of injected muons, $\sigma_{\mu I \to N X}$ is the cross section in \cref{fig:energy_dep_production_xs}, $E_\mu(\ell)$ is the distance-dependent muon energy computed in \cref{eq:muon_range} for traversed distance $\ell \equiv X/\rho_\oplus$, and $n_\oplus$ is the nucleon number density in the Earth.
\cref{fig:earth_production_pos} shows the distribution of HNL production locations along the terrestrial beamline for different HNL masses, as well as those for neutrino backgrounds from muon decay and scattering.

The momentum of outgoing HNLs produced in terrestrial upscattering is fixed by sampling differential cross sections computed in \texttt{MadGraph5\_aMC@NLO} for an isoscalar target.
The direction of the primary muon is offset from the beam direction due to multiple Coulomb scattering in the Earth.
The RMS angle is given by~\cite{ParticleDataGroup:2024cfk}
\begin{equation} \label{eq:muon_mcs}
\theta_{\rm RMS} = \frac{13.6~{\rm MeV}}{\beta c p_\mu}\sqrt{X/X_0},
\end{equation}
where $X$ is the traversed column depth and $X_0 = 26.5$ is the radiation length of rock.
The accumulated deviation angle of muons in our setup must be found by integrating the variance of the angle \cref{eq:muon_mcs} over the traversed column depth, accounting for energy loss over the path,
\begin{equation}
\theta_{\rm RMS}^{\rm tot} = \sqrt{\int_0^{\ell_{\rm max}} [\theta_{\rm RMS}(E_\mu(\ell))]^2 d\ell}.
\end{equation}
This smearing of the initial muon direction, which amounts to approximately $0.5~{\rm mrad}$ over one kilometer of Earth for a $5~{\rm TeV}$ muon, is propagated to the outgoing HNL direction in our simulation.

\subsubsection{HNL Detection in the Atmosphere}

The produced HNLs travel a mean distance $\ell_d = \gamma\beta c/\Gamma_N$ before decaying; for the relevant masses and couplings $\ell_d$ ranges from tens of kilometers up to and beyond the available atmospheric baseline, so that a fraction $1 - e^{-d_{\rm max}/\ell_d}$ of HNLs decay within the detectable range of an air Cherenkov camera with atmospheric baseline $d_{\max}$.
As shown in \cref{fig:atmo_dump_geometry}, $d_{\rm max}$ can range anywhere from $\mathcal{O}(1)~{\rm km}$ to $\mathcal{O}(100)~{\rm km}$, depending on the depth and angle of the beam dump.

The visible signal is the Cherenkov light radiated by the charged decay products as they propagate through the atmosphere.
During the sufficiently dark period of the night, this Cherenkov light can be observed by optical camera detectors~\cite{Otte:2018uxj}.
For both HNL decays and neutrino interactions in the atmosphere, these can include long muon tracks and/or hadronic showers.
The number of Cherenkov photons radiated by a charged particle per unit length is given by
\begin{equation} \label{eq:frank_tamm}
\frac{dN}{dx} = 2 \pi \alpha z^2 \sin^2\theta_C \bigg[ \frac{1}{\lambda_{\rm min}} - \frac{1}{\lambda_{\rm max}}\bigg],
\end{equation}
where $z=1$ is the charge, $\theta_C \approx 1.4^\circ$ is the Cherenkov angle in the atmosphere, and $\lambda_{\rm min/max}$ are the bounds on the observable photon wavelength, taken to be $300~{\rm nm}$ and $1000~{\rm nm}$, respectively.
A relativistic muon radiates approximately $64$ photons per meter of track in air over the $300$--$1000$~nm band ($n_{\rm air}\approx1.0003$ at sea level).
Hadronic final states instead produce extended showers, whose Cherenkov yield we evaluate with a track-length integral.
The electromagnetic component of a shower of energy $E$ has total charged-particle path length $X_0/\rho \,(E/E_c)$, with $X_0 = 36.6~{\rm g\,cm^{-2}}$ the radiation length and $E_c \approx 80~{\rm MeV}$ the critical energy in air; converting to a physical length and
folding in \cref{eq:frank_tamm} gives $C_{\rm CH}\approx1.9\times10^{4}$ photons per unit $E/E_c$.
This product is to good approximation altitude independent, since the per-unit-length yield scales with the air density $\rho$ while the track length scales as $1/\rho$.

Only particles above the Cherenkov threshold $E_{\rm thr}(h) = m_ec^2/\sqrt{2\delta(h)}$ actually
radiate, however, whereas the track length above counts the entire charged path down to zero
energy.
We therefore weight the integral by the radiating fraction of the cascade, obtained by
integrating the universal electron energy spectrum of Ref.~\cite{Nerling:2005fj} from
$\ln E_{\rm thr}$ upward with the Cherenkov weight $1-(E_{\rm thr}/E)^2$.
This gives $W = 0.46$ at sea level ($E_{\rm thr}=21~{\rm MeV}$) and $W = 0.40$ at $5~{\rm km}$ ($E_{\rm thr}=28~{\rm MeV}$); omitting it would overestimate the yield by a factor $\approx2.2$.
Since $E_{\rm thr}\propto\rho^{-1/2}$, this restores a weak altitude dependence, but it amounts to $<15\%$ over the $0$--$5~{\rm km}$ range relevant here.
The total yield is then
\begin{equation}
  N_\gamma = f_{\rm EM}(E)\,\frac{E}{E_c}\,C_{\rm CH}\,W(h),
\end{equation}
with the electromagnetic fraction $f_{\rm EM}(E) = 1-(E/{\rm GeV})^{-0.14}$ following the standard hadron-calorimetry parametrization~\cite{Groom:2006am}.

The shower photons are spread over an angle much larger than the Cherenkov angle in air
$\theta_C\approx1.4^\circ$, so their directions trace the angular distribution of the shower rather
than the Cherenkov cone.  We model this as a Gaussian in solid angle whose width is matched to the
RMS of the Cherenkov angular distribution of Ref.~\cite{Nerling:2005fj}, giving
$\sigma_{\rm had} = 7.0^\circ$ at sea level and $5.9^\circ$ at $5~{\rm km}$.  That parametrization
is derived for ultra-high-energy air showers; we evaluate it at shower maximum, where the angular
spread is set by the local air density through $E_{\rm thr}$ rather than by the primary energy.
The $N\to\nu_\mu\,\mu^+\mu^-$ signal is unaffected, its final state being purely leptonic and producing no hadronic shower --- $\sigma_{\rm had}$ enters only through the hadronic veto applied to the charm background in the next section.
Photons en route to the detector are attenuated by Rayleigh ($\propto\lambda^{-4}$, scale height $\approx 8.5~{\rm km}$) and aerosol Mie ($\propto\lambda^{-1.2}$, scale height $\approx 1.2~{\rm km}$) scattering.
\Cref{sec:cherenkov_attenuation} provides more details on our treatment of atmospheric attenuation.
The detector is a TRINITY-style SiPM camera of radius $R_{\rm det} = 4~{\rm m}$ and angular resolution $\sim 0.3^\circ$ per pixel; folding in a photon detection efficiency of $0.4$ and the per-pixel night-sky background, an charged particle is registered when the clustered signal exceeds $\sim 12$ photoelectrons ($\approx 30$ detected photons)~\cite{otte2018studies-e8c}.


\subsubsection{Search Strategy and Background Mitigation}

\begin{figure}
    \centering
    \includegraphics[width=\linewidth]{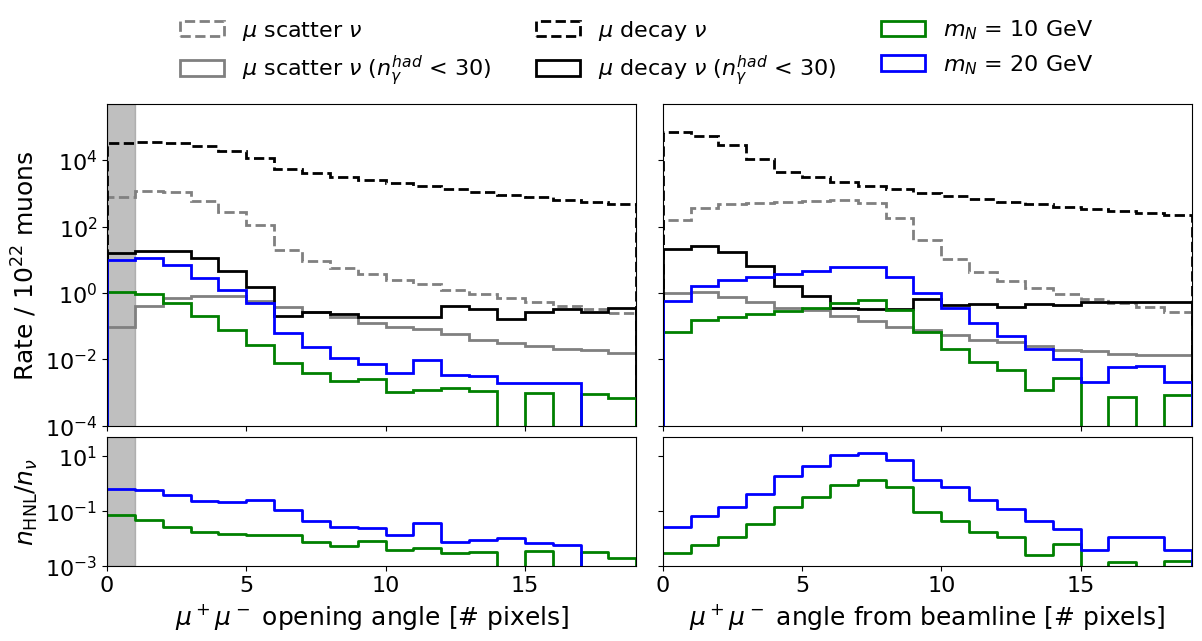}
    \caption{Signal and background distributions for two dimuon observables: the opening angle $\Delta \phi_{\mu \mu}$ (left) and the angle between with respect to the beamline $\theta_{\mu \mu}$ (right).
    Distributions are shown in units of the $0.3^\circ$ camera pixels.
    The blue and green lines correspond to two different HNL masses, each at $|U_{\mu N}|^2 = 10^{-12}$.
    The black and gray lines correspond to dimuon backgrounds from neutrinos produced in muon decay and scattering, respectively.
    The dashed (solid) lines indicate the background rate before (after) removing charm-producing neutrino interactions with visible hadronic showers.
    The bottom panels show the signal to background ratio.
    The vertical gray band in the left figure indicates that dimuons registering on the same pixel are removed from the signal definition.} 
    \label{fig:dimuon_observables}
\end{figure}

For this configuration, we take our signal definition to be HNL decays to two muons, $N \to \nu_\mu\,\mu^+\mu^-$.
The dimuon signature eliminates most of the Standard Model background at the cost of the dimuon branching ratio, which opens above $m_N = 2m_\mu$ and grows to the $\sim 10\%$ level at higher masses as shown in \cref{fig:BRmuDump}.
Like the dense-target configuration of \cref{sec:dump}, there are backgrounds from neutrinos produced by the scattering and decay of muons travsering the target bedrock.
Charged-current $\nu_\mu$ interactions in the atmosphere that produce charm hadrons can result in dimuon final states.
These neutrino-induced dimuons constitute a background to the HNL signal when the hadronic shower is below detection threshold.

After HNL momenta are sampled according to the procedure in \cref{sec:balloon_hnl_production}, we use the \texttt{SIREN} simulation package~\cite{Schneider:2024eej} to sample outgoing dimuon momenta from the differential decay width.
Neutrino dimuon backgrounds follow a similar pipeline; after neutrinos are generated from muon scattering and decay in the Earth, their charm-producing interactions in the atmosphere are simulated with \texttt{SIREN}.
The charm production cross section is computed using the EPPS21NLO nuclear parton distribution functions for an oxygen target~\cite{Eskola:2021nhw}, accounting for fragmentation into $D^0$, $~\bar{D}^0$, $D^\pm$, and $D_s^\pm$ final states.
Outgoing charm hadron kinematics are sampled from the Peterson parameterization of the fragmentation function~\cite{Peterson:1982ak}, and muonic charm decays are sampled according to their branching ratios from the Particle Data Group~\cite{ParticleDataGroup:2026mpi} following three body decay kinematics.
For both signal and background we use a Monte Carlo algorithm for Cherenkov light generation and propagation through the atmosphere, recording the number of Cherenkov photons reaching the circular Cherenkov camera for each muon track or hadronic shower.

Atmospheric HNL decays are recorded as signal if both muons result in at least 12 photoelectrons in different pixels of the camera detector.
Neutrino dimuon backgrounds survive if they satisfy the same criteria, with the added condition that the hadronic shower produces less than 12 photoelectrons in the camera (i.e., it mimics the dimuon-only signature of HNLs).
We consider two observables to separate HNL signal from surviving neutrino backgrounds: the opening angle between the dimuon pair $\Delta \phi_{\mu \mu}$, and the angle between the dimuon system and the beamline $\theta_{\mu \mu}$.
These are shown in \cref{fig:dimuon_observables}.
Dimuons from HNL decay tend to be more collimated and further separated from the beamline compared to neutrino backgrounds, the latter being a direct consequence of the angular distributions shown in \cref{fig:E_theta_distribution}.

\subsubsection{Sensitivity}

\begin{figure}
    \centering
    \includegraphics[width=\linewidth]{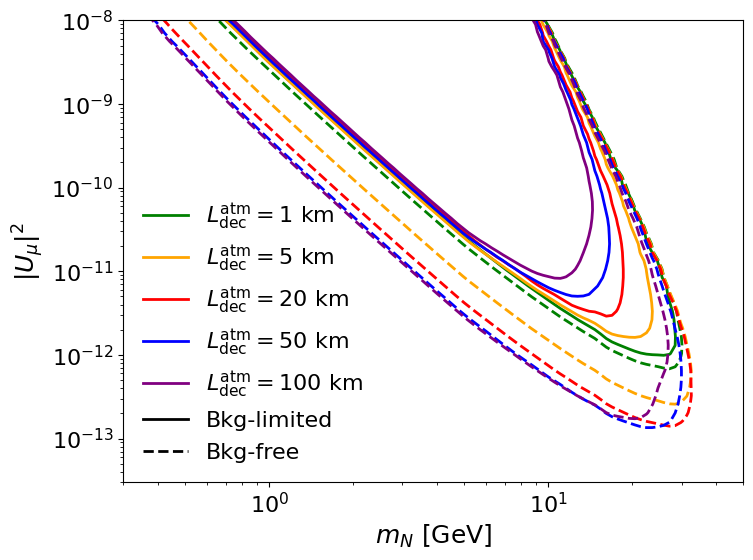}
    \caption{The background-limited (solid) and background-free (dashed) $2\sigma$ sensitivity of the atmospheric beam dump setup, for five different atmospheric decay region lengths.
    The background-limited sensitivities account for neutrino-induced dimuon background via an $S/\sqrt{S +B}$ test statistic.}
    \label{fig:balloon_distance_tradeoff}
\end{figure}

To compute sensitivity, we first make hard cut requiring $0.3^\circ < \Delta \phi_{\mu \mu} < 5^\circ$.
We then calculate a binned $S/\sqrt{S+B}$ test statistic in $\theta_{\mu \mu}$.
The ability to separate signal from background depends most strongly on the efficiency of rejecting visible hadronic showers in neutrino-induced dimuon backgrounds.
This efficiency increases at shorter atmospheric distances such that, somewhat counterintuitively, the most sensitive setup features an atmospheric decay region length of approximately $1~{\rm km}$.
This can be achieved with a Cherenkov camera at a modest altitude of $50~{\rm m}$, such that the camera could be placed on a small hill in an otherwise flat area.
In \cref{fig:balloon_distance_tradeoff}, we show the background-limited and background-free sensitivity over a range of atmospheric decay lengths.
As expected, without background the sensitivity improves with increasing atmospheric decay region length up to $20~{\rm km}$, while the background-limited sensitivity is best for a $1~{\rm km}$ decay region.

\begin{figure}
    \centering
    \includegraphics[width=\linewidth]{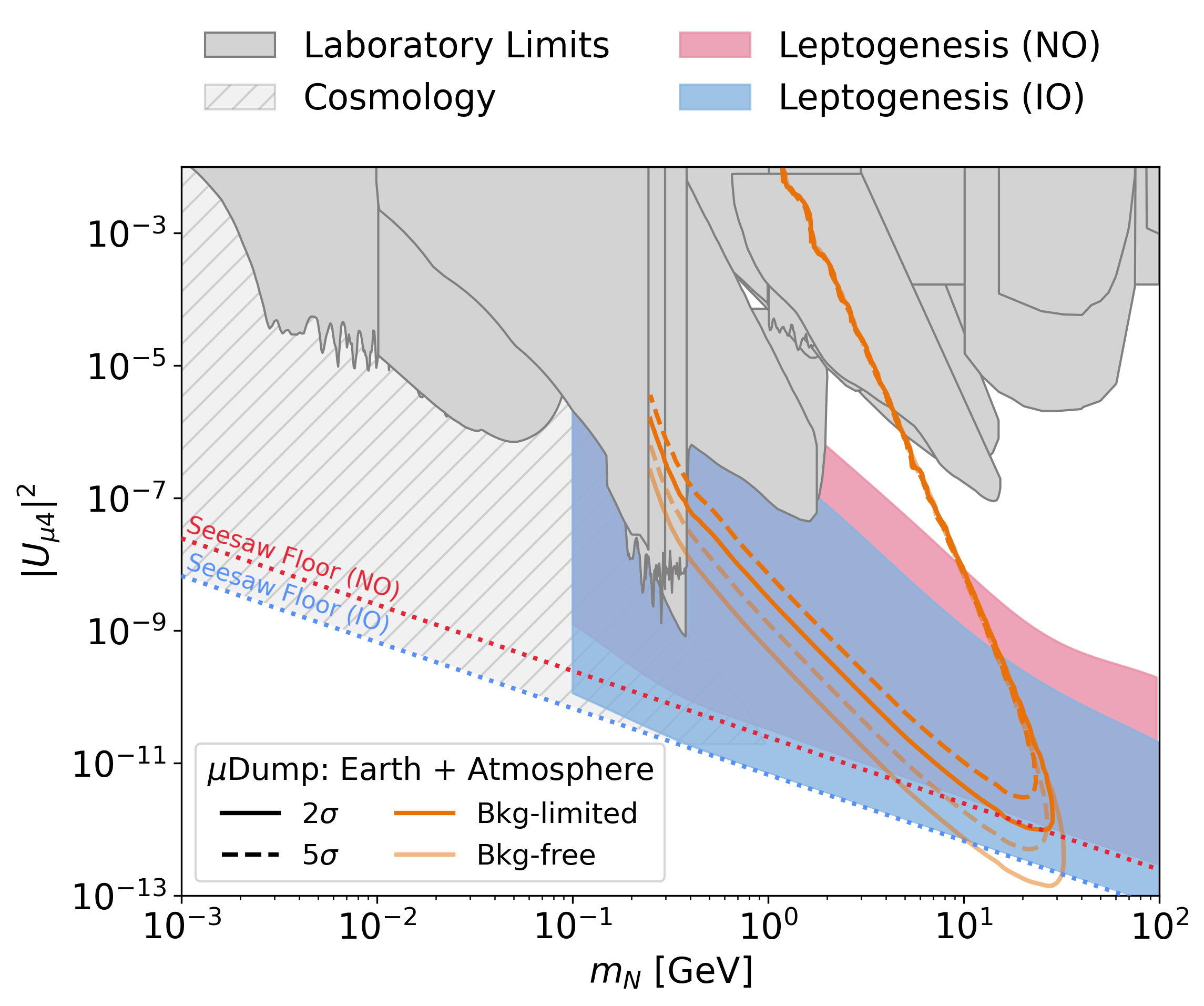}
    \caption{Sensitivity of the atmospheric beam-dump configuration in $(m_N,\, \UNsq{\mu})$ parameter space at the $2\sigma$ (solid) and $5\sigma$ (dashed) confidence level.
    Background-limited sensitivity considers a $1~{\rm km}$ and incorporates neutrino-induced dimuon backgrounds, while the background-free sensitivity considers a $20~{\rm km}$ decay region and represents the maximum sensitivity of this configuration without neutrino-induced backgrounds.
    Gray regions denote existing limits in the parameter space from Ref.~\cite{Fernandez-Martinez:2023phj}, including BBN constraints~\cite{Sabti:2020yrt} represented by the hatched gray region.
    Red and blue regions denote the parameter space preferred by the low-scale leptogenesis models discussed in \cref{sec:leptogenesis}, while the dotted lines represent the mixing floor predicted by the seesaw mechanism.
}
    \label{fig:balloon_sens}
\end{figure}

\Cref{fig:balloon_sens} summarizes the projected sensitivity of the atmospheric beam dump in the $(m_N,\, \UNsq{\mu})$, assuming $N_\mu = 10^{22}$ muons on target.
Following \cref{fig:balloon_distance_tradeoff}, the background-limited case assumes a $1~{\rm km}$ decay region, while the background-free case assumes a $20~{\rm km}$ region, representing the maximum reach of an HNL search with better rejection of neutrino backgrounds.
The reach extends to small mixing across the $m_N \sim$~few--tens of GeV range, and with improved neutrino background rejection this setup would begin to probe the theoretically-motivated seesaw and leptogenesis bands discussed in \cref{sec:leptogenesis}.
The required background rejection to reach the Seeesaw line could potentially be achieved by a multi-camera setup to better tag hadronic showers, or higher-altitude free-flying or satellite-borne platforms that leverage the expoenentially-decaying atmoshperic density to minimize neutrino scattering backgrounds.
We leave a detailed study of these more complicated detector arrangements to future work.

\FloatBarrier
\subsection{Earth Target Plus Vacuum Decay Region} \label{sec:terrestrial_vacuum}

In this subsection, we combine the most powerful aspects of the configurations in \cref{sec:dump} and \cref{sec:balloon} into an HNL search with the capability to reach the seesaw floor.
For HNL production we consider the terrestrial target of \cref{sec:balloon}, in which the muon beam is directed through $1~{\rm km}$ of bedrock, upscattering into HNLs according to the cross section in  \cref{fig:energy_dep_production_xs}.
We then use the vacuum decay region proposed in \cref{sec:dump} to look for HNL decays without any backgrounds from neutrino interactions in the detector.

\begin{figure}
    \centering
    \includegraphics[width=\linewidth]{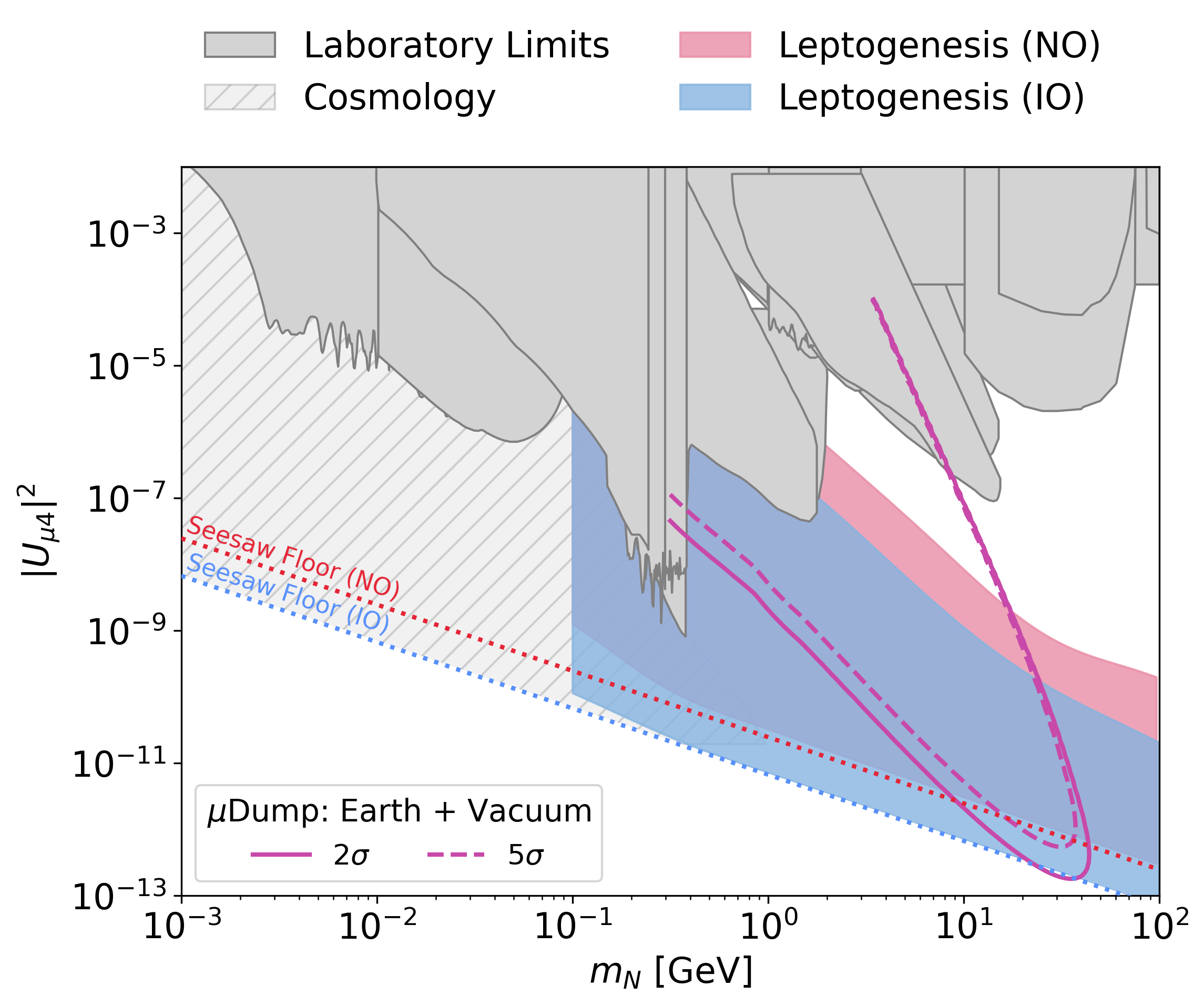}
    \caption{Sensitivity of the terrestrial-plus-vacuum beam-dump configuration in $(m_N,\, \UNsq{\mu})$ parameter space at the $2\sigma$ (solid) and $5\sigma$ (dashed) confidence level.
    Contours are based on a $100~{\rm m}$ long decay region with pressure low enough to restrict neutrino-induced backgrounds to less than one event in $10^{22}$ muons on target.
    }
    \label{fig:terrestrial}
\end{figure}

We follow the same procedure from the earlier sections, using \cref{eq:terrestrial_HNL_production} to compute the HNL production rate.
HNLs are generated along the $1~{\rm km}$ muon beamline and tracked to their intersections with the $100~{\rm m}$ long, $5~{\rm m}$ radius vacuum decay region.
The $2\sigma$ $(5\sigma)$ sensitivity of this setup is shown in \cref{fig:terrestrial}, determined by requiring approximately 3.8 (15) HNL decays to any final state in the detector.

As in the previous beam-dump configurations, there are backgrounds from neutrinos produced in muon decay and charged-current scattering as the muons traverse the terrestrial beam dump.
In order to observe less than $0.1$ neutrino interaction in the decay region, the entire tunnel must be held at a pressure of
\begin{equation} \label{eq:terrestrial_pressure_req}
P \lesssim 3.6 \times 10^{-6}~{\rm mbar}.
\end{equation}
This requirement is stricter than \cref{eq:vacP} because there are more neutrinos produced in the terrestrial beam dump compared to the tungsten or water target of \cref{sec:dump}.
Still, it is higher than the $10^{-6}~{\rm mbar}$ pressure demonstrated by the vacuum decay region of NA62~\cite{NA62:2023qyn}.
This requirement could be loosened to $10^{-4}~{\rm mbar}$ by rejecting events within approximately $10~{\rm cm}$ of the beam axis, where the bulk of the neutrino backgrounds from muon decay are expected to occur, as shown in \cref{fig:E_theta_distribution}.

As shown in \cref{fig:terrestrial}, this terrestrial-plus-vacuum beam-dump configuration can reach the seesaw floor in $|U_{\mu N}|^2$ the inverted ordering for HNL masses $m_N \sim 30~{\rm GeV}$.
To our knowledge, the muon beam-dump proposal is the first to reach the seesaw floor in any flavor for HNL masses $m_N > 1~{\rm GeV}$.
It also covers much of the parameter space favored by ARS models of leptogenesis, as discussed in \cref{sec:leptogenesis}.
Thus, through a relatively straightforward beam dump extension, a future muon collider would take a significant step toward understanding the origin of the nonzero neutrino masses discovered by oscillation experiments at the beginning of the millennium.

\section{Muon-ion collider} \label{sec:muon_ion}

\begin{figure}
    \centering
    \includegraphics[width=\linewidth]{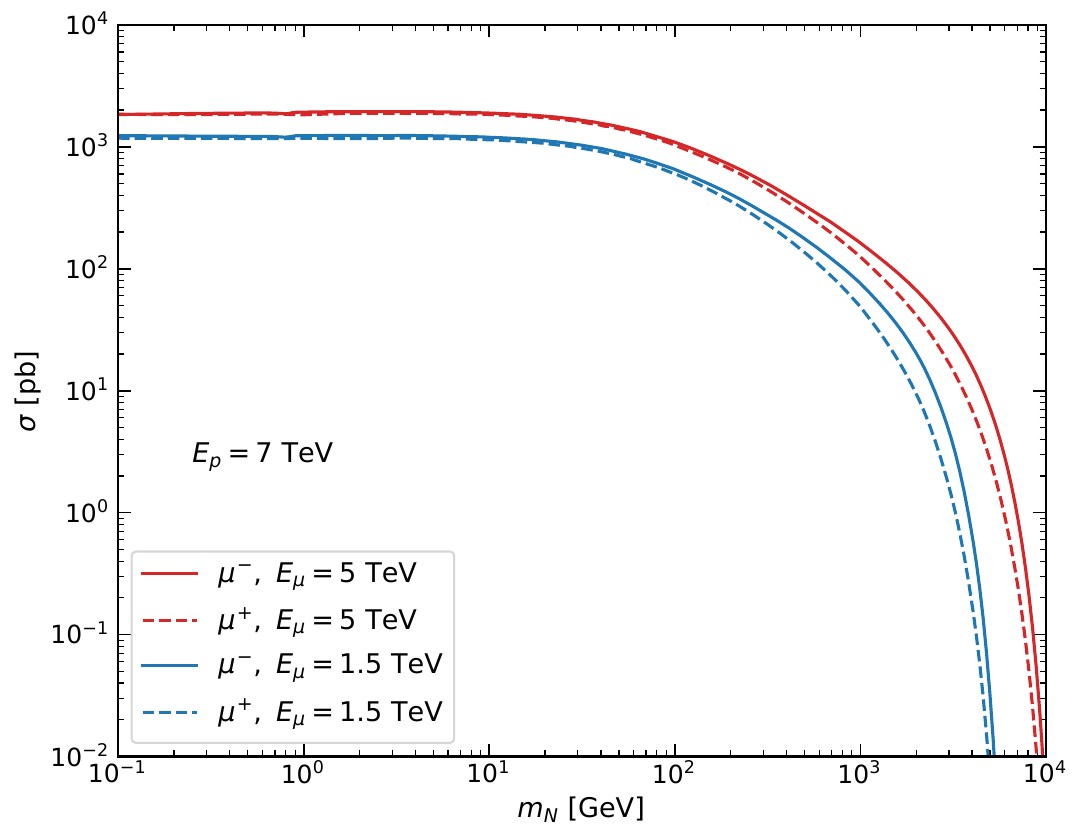}
    \caption{HNL production cross section $\sigma(\mu^\pm p \to N X)$ as a function of $m_N$ for $\mu^-$ and $\mu^+$ beams at $E_\mu = 1.5~\tev$ and $5~\tev$ colliding with a $7~\tev$ proton beam~\cite{Akturk:2025ubm}, normalized to $\UNsq{\mu} = 1$. Calculations are performed with \texttt{MadGraph5\_aMC@NLO} using the CT18NNLO PDF set~\cite{Hou:2019efy}.}
    \label{fig:XSMuIC}
\end{figure}

A muon-ion collider, in which one muon beam of the muon collider is brought into collision with a $7~\tev$ proton beam from the (HL-)LHC~\cite{Acosta:2022ejc,Akturk:2025ubm}, provides a complementary avenue for HNL searches.
The production mechanism is the same deep-inelastic scattering process $\mu^{\pm} p \to N X$ via $t$-channel $W$-boson exchange as in the beam dump, now operating in collider mode at a center-of-mass energy $\sqrt{s} = 2\sqrt{E_\mu E_p} \approx 6.5~\tev$ ($11.8~\tev$) for $E_\mu = 1.5~\tev$ ($5~\tev$).
The HNL production cross section as a function of $m_N$ is shown in \cref{fig:XSMuIC}.

\begin{figure}
    \centering
    \includegraphics[width=\linewidth]{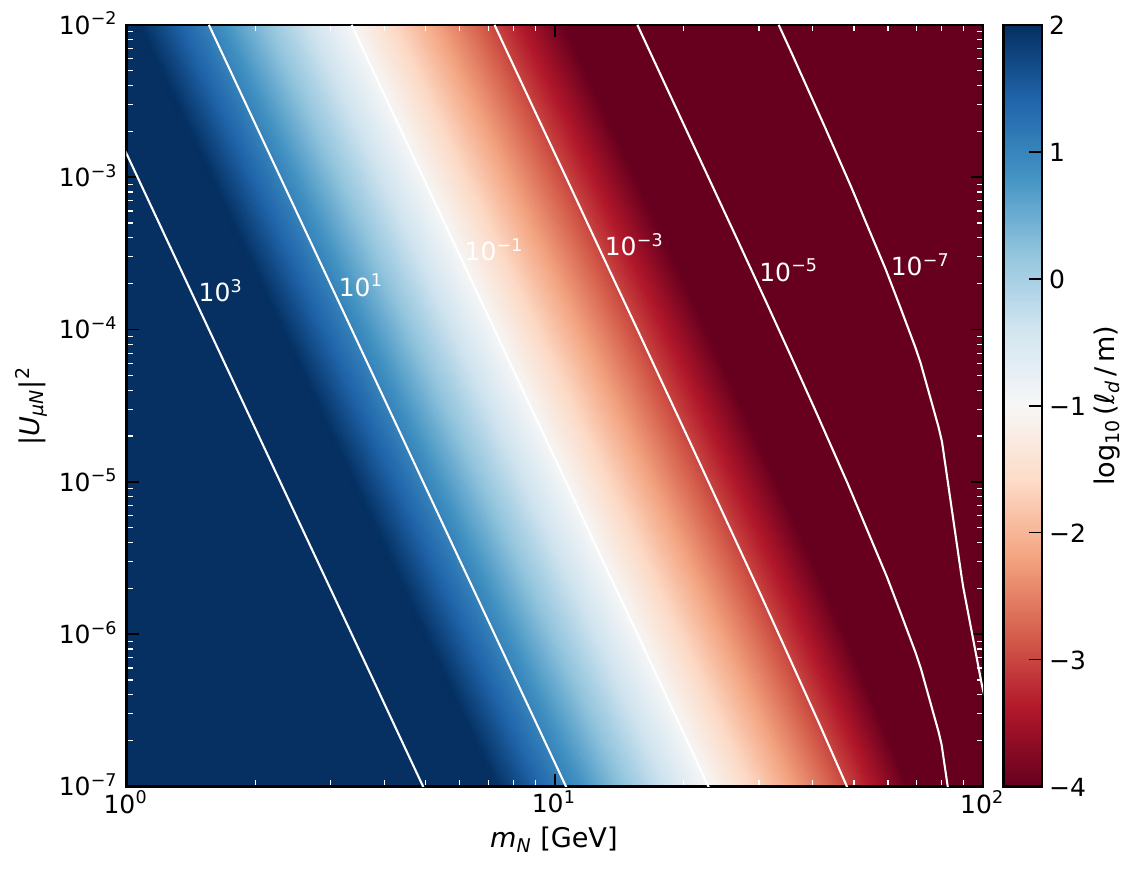}
    \caption{Mean HNL decay length $\log_{10}(\ell_d/\mathrm{m})$ in the $(m_N,\,\UNsq{\mu})$ plane in the muon-ion collider lab frame at $E_\mu = 5~\tev$. White iso-contour lines are labeled with $\ell_d$ values in meters.}
    \label{fig:decay_length}
\end{figure}

In the collider environment, produced HNLs are highly boosted and travel a macroscopic distance before decaying.
The HNL decay length $\ell_d$ is shown in \cref{fig:decay_length} as a function of $(m_N,\,\UNsq{\mu})$ for $E_\mu = 5~\tev$.

We require a displaced decay vertex in the range $L_0 = 0.4~\mathrm{mm} < d < L_1 = 11~\mathrm{m}$, spanning the inner tracker to the outer muon system of a general-purpose collider detector, as adopted in long-lived-particle searches at the LHC and the proposed FCC-ee~\cite{Alimena:2019zri,Blondel:2022qqo}.
We assume integrated luminosities of $1~\mathrm{ab}^{-1}$ at $1.5~\tev$ and $10~\mathrm{ab}^{-1}$ at $5~\tev$~\cite{Akturk:2025ubm}.
The displaced-vertex requirement, together with the comparatively clean collider environment, strongly suppresses prompt Standard Model backgrounds, and we therefore treat the search as background-free, as in analogous long-lived-particle searches~\cite{Alimena:2019zri,Blondel:2022qqo,Bellagamba:2025xpd}.
The number of HNL decays reconstructed within the fiducial volume is
\begin{equation}\label{eq:Ndec}
    N_{\rm dec} = \sigma_0(m_N)\,\UNsq{\mu}\,\mathcal{L}\,
    \left( e^{-L_0/\ell_d} - e^{-L_1/\ell_d} \right),
\end{equation}
where $\sigma_0(m_N) \equiv \sigma(m_N)$ is the production cross section normalized to unit mixing, shown in \cref{fig:XSMuIC}, and the factor in parentheses is the probability for the HNL to decay within the fiducial volume.
We adopt an idealized detector and assume unit signal efficiency  for reconstructing such displaced vertices. A realistic efficiency would weaken the quoted reach by an $\mathcal{O}(1)$ factor.
Since the search is background-free, the $2\sigma$ ($5\sigma$) sensitivity contour corresponds to $N_{\rm dec} \simeq 3.8$ ($\simeq 15$) decays.

\begin{figure}
    \centering
    \includegraphics[width=\linewidth]{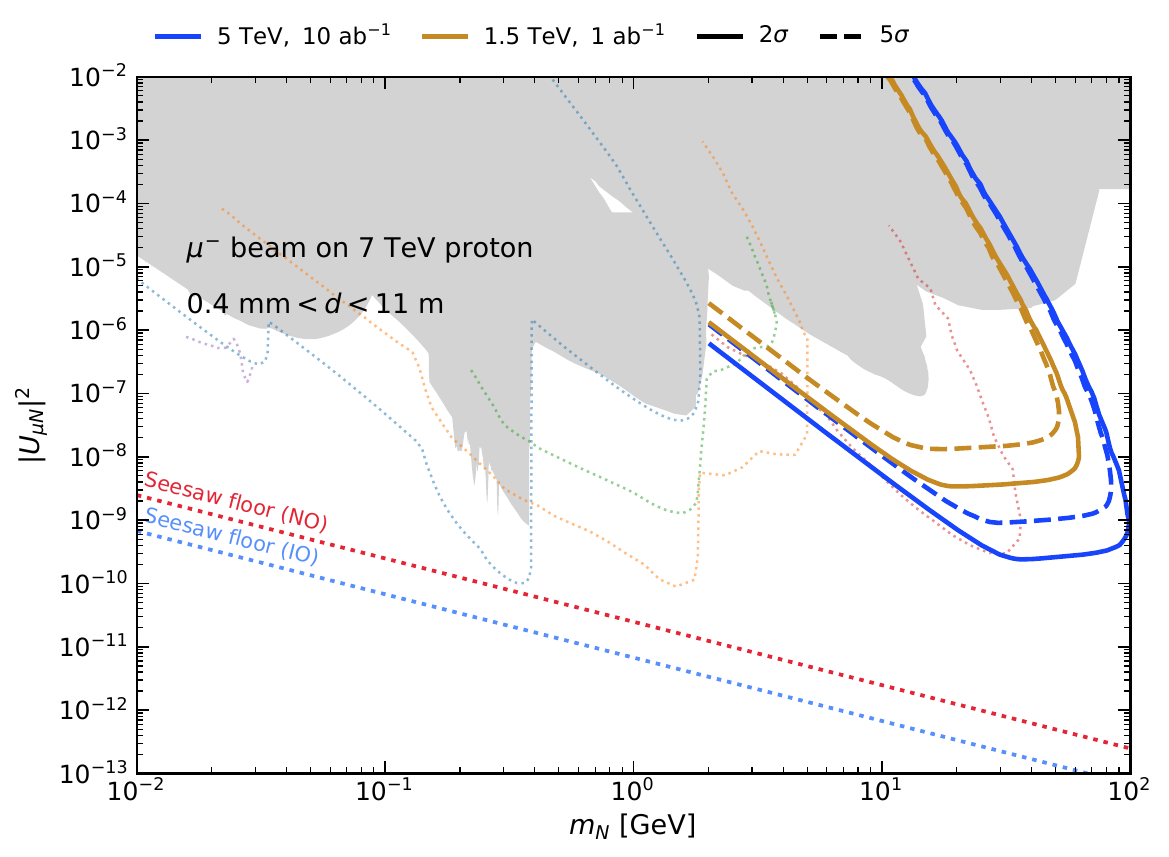}
    \caption{Projected $2\sigma$ (solid) and $5\sigma$ (dashed) sensitivity in the $(m_N,\,\UNsq{\mu})$ plane from the muon-ion collider with $E_\mu = 1.5~\tev$ at $\mathcal{L} = 1~\mathrm{ab}^{-1}$ (amber) and $E_\mu = 5~\tev$ at $\mathcal{L} = 10~\mathrm{ab}^{-1}$ (blue). Only the $\mu^-$ beam is shown.}
    \label{fig:MuICSens}
\end{figure}

Because the decay products of light HNLs are too soft to satisfy realistic trigger and vertex-reconstruction requirements, we restrict the reach to $m_N \gtrsim 2~\gev$, following the FCC-ee analyses~\cite{Blondel:2022qqo,Bellagamba:2025xpd}.
The projected $2\sigma$ (solid) and $5\sigma$ (dashed) sensitivities in the $(m_N,\,\UNsq{\mu})$ plane are shown in \cref{fig:MuICSens} for the $\mu^-$ beam.
At $E_\mu = 1.5~\tev$ with $1~\mathrm{ab}^{-1}$, the muon-ion collider probes $\UNsq{\mu}$ down to $\sim 3\times10^{-9}$ ($\sim 1\times10^{-8}$) at $2\sigma$ ($5\sigma$), near $m_N \sim 15$--$20~\gev$.
The $5~\tev$ configuration with $10~\mathrm{ab}^{-1}$ deepens the reach to $\UNsq{\mu} \sim 2\times10^{-10}$ ($\sim 9\times10^{-10}$) at $2\sigma$ ($5\sigma$), near $m_N \sim 30$--$35~\gev$, and extends sensitivity to HNL masses of several tens of GeV.

\begin{figure}
    \centering
    \includegraphics[width=\linewidth]{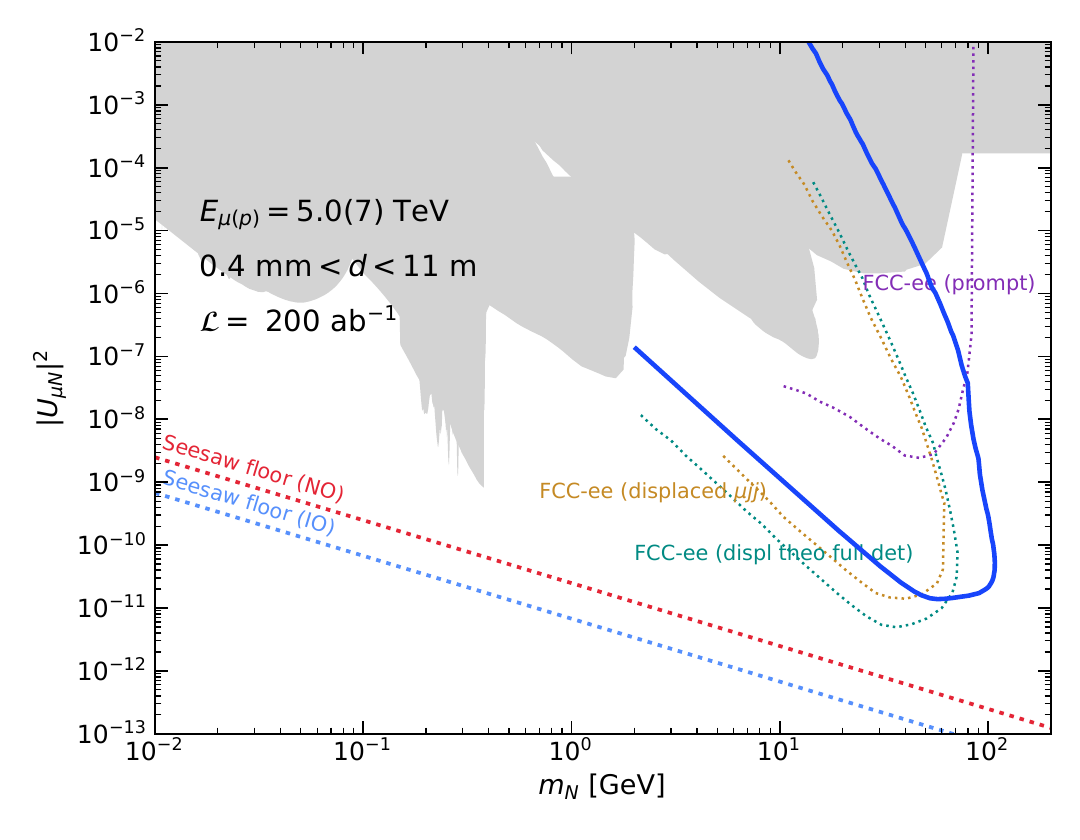}
    \caption{Comparison of the muon-ion collider $2\sigma$ projected sensitivity  at $E_{\mu} = 5~\tev$ with FCC-ee projections~\cite{Bellagamba:2025xpd,Blondel:2022qqo}.}
    \label{fig:MuICFCC}
\end{figure}

To assess the ultimate potential of the muon-ion collider, we compare its reach with the projected sensitivity of the FCC-ee in \cref{fig:MuICFCC}, adopting an optimistic integrated luminosity of $\mathcal{L} = 200~\mathrm{ab}^{-1}$ at $E_\mu = 5~\tev$ to match the FCC-ee exposure for a like-for-like comparison.

With this exposure, the muon-ion collider probes $\UNsq{\mu}$ down to $\sim 1.4\times10^{-11}$ near $m_N \sim 56~\gev$, comparable to the FCC-ee displaced $\mu jj$ projection but about a factor of three above the deepest reach of the FCC-ee full-detector displaced-vertex search ($\sim 5\times10^{-12}$).
Across the overlapping mass range the FCC-ee searches are generally more sensitive, particularly at lower masses.
The muon-ion collider is instead complementary in its mass coverage: by virtue of the higher center-of-mass energy it remains sensitive out to $m_N \sim 100~\gev$ and beyond, past the kinematic reach of the FCC-ee searches. Throughout, we show the $\mu^-$ beam as representative, and the $\mu^+$ configuration yields a comparable reach.

\section{Leptogenesis targets}
\label{sec:leptogenesis}
The same GeV-scale HNLs probed by the neutrino factory searches described above provide a minimal mechanism for generating the observed baryon asymmetry of the universe through Akhmedov--Rubakov--Smirnov (ARS) leptogenesis~\cite{Akhmedov:1998qx,Asaka:2005pn}.
In this scenario, CP-violating oscillations among nearly degenerate HNLs produce flavor-dependent lepton asymmetries in the thermal plasma during their freeze-in production, which are partially converted into a baryon asymmetry by electroweak sphalerons before these decouple at $T_{\rm sph} \simeq 131.7~\gev$.
The asymmetry must be generated before sphaleron freeze-out, many orders of magnitude in time before BBN.
We consider HNL masses $0.1~\gev \leq m_N \leq 100~\gev$, the mass window targeted by the neutrino factory.
It is therefore important to identify the region of the $(m_N, |U_{\alpha N}|^2)$ parameter space in which successful baryogenesis can be achieved, and to assess whether the proposed experimental configurations can probe this theoretically motivated target.
We note that the same region is of interest beyond leptogenesis: GeV-scale right-handed neutrinos arise in a variety of low-scale seesaw constructions~\cite{Abdullahi:2022jlv}, for instance in models where lepton number is broken spontaneously and the seesaw scale is tied to the decay constant of the associated Majoron~\cite{Chikashige:1980ui,Gelmini:1980re}, with implications for cosmology and the cosmic neutrino background~\cite{Cline:2026tkp}.

\subsection{Seesaw parametrization and the seesaw floor}
\label{sec:seesaw_floor}

We work in the minimal type-I seesaw framework with $n=2$ right-handed neutrinos $N_2$ and $N_3$, so that the lightest active neutrino is exactly massless ($m_1 = 0$ for normal ordering, NO; $m_3 = 0$ for inverted ordering, IO).
After electroweak symmetry breaking the Dirac mass is $m_D = Y v$ with $v \simeq 174~\gev$, the light neutrino mass matrix is
\begin{equation}
m_\nu \simeq -m_D \hat{M}_N^{-1} m_D^{T},
\end{equation}
and the heavy states acquire a small admixture of the active flavors, $\Theta = m_D \hat{M}_N^{-1}$, with $|U_{\alpha N_I}|^2 \equiv |\Theta_{\alpha I}|^2$.
The active--heavy mixing is parametrized through the Casas--Ibarra construction~\cite{Casas:2001sr},
\begin{align}
    \Theta &= i\, U_{\rm PMNS}\, \sqrt{\hat{m}_\nu}\, R^{T}\, \hat{M}_N^{-1/2}\,,\nonumber\\
    Y &= \frac{i}{v}\, U_{\rm PMNS}\, \sqrt{\hat{m}_\nu}\, R^{T}\, \hat{M}_N^{1/2}\,,
    \label{eq:CI}
\end{align}
where $\hat{m}_\nu = \text{diag}(m_1, m_2, m_3)$ contains the light neutrino masses, $\hat{M}_N = \text{diag}(M_2, M_3)$ the heavy masses, and $R$ is a complex orthogonal matrix ($RR^{T}=\openone$).
Note that the mixing $\Theta$ is independent of $v$; the vacuum expectation value enters only the Yukawa coupling, $Y = \Theta \hat{M}_N / v$, which controls the HNL production rate in the early universe.
For $n=2$, $R$ contains a single complex angle $\omega = \xi + i\eta$ that must rotate in the plane of the two massive light states, i.e.\ $(m_2,m_3)$ for NO and $(m_1,m_2)$ for IO; placing the rotation in a plane containing the massless state yields a rank-one Yukawa that reproduces only one of the two nonzero light masses~\cite{Drewes:2016jae}. We use NuFIT~5.3 best-fit oscillation parameters~\cite{Esteban:2020cvm} throughout.

\paragraph{Summed mixing.}
Summing $|\Theta_{\alpha I}|^2$ over flavors and over the two quasi-degenerate heavy states, PMNS unitarity removes interference terms and one finds the relation~\cite{Drewes:2016jae,Drewes:2019mhg}
\begin{equation}
    U^2 \equiv \sum_{\alpha,I} |U_{\alpha N_I}|^2 = \frac{\sum_i m_i}{m_N}\,\cosh 2\eta\,,
    \label{eq:U2_CI}
\end{equation}
with $m_N \equiv \tfrac12(M_2+M_3)$, up to corrections of order $\delta M/m_N$.
Since $\cosh 2\eta \geq 1$, the total mixing is bounded from below by the seesaw floor
\begin{equation}
    U^2_{\rm min} = \frac{\sum_i m_i}{m_N}
    = \begin{cases}
        (m_2 + m_3)/m_N \simeq 0.059~\text{eV}/m_N & \text{(NO)}\\[2pt]
        (m_1 + m_2)/m_N \simeq 0.099~\text{eV}/m_N & \text{(IO)}\,,
    \end{cases}
    \label{eq:seesaw_floor}
\end{equation}
attained at $\eta = 0$.
Within this minimal model no choice of parameters reproduces the measured mass splittings below this line; the floor is a property of the seesaw alone and holds independently of whether the HNLs generate the baryon asymmetry.
Values of $U^2$ far above the floor require $|\eta|\gg 1$, i.e.\ large cancellations in $m_\nu$ characteristic of an approximate lepton-number symmetry.

\paragraph{Per-flavor mixings.}
Individual-flavor lower bounds are weaker and depend on the PMNS parameters through interference.
For a single flavor $\alpha$ the interference between the two massive light states survives,
\begin{equation}
\begin{split}
    |U_{\alpha N}|^2 &\equiv \sum_I |U_{\alpha N_I}|^2 \\
    &= \frac{1}{m_N}\Big[\big(m_a |U_{\alpha a}|^2 + m_b |U_{\alpha b}|^2\big)\cosh 2\eta \\
    &\qquad + 2\sqrt{m_a m_b}\,\mathrm{Im}\big(U_{\alpha a}U_{\alpha b}^*\big)\sinh 2\eta\Big]\,,
\end{split}
    \label{eq:Ualpha2}
\end{equation}
where $(a,b)$ label the two massive states and $U_{\alpha i}$ are PMNS elements (see also Ref.~\cite{Drewes:2016jae}).
The second term can be negative, so a single flavor can be driven below the incoherent sum $(m_a|U_{\alpha a}|^2+m_b|U_{\alpha b}|^2)/m_N$; minimizing over $\eta$ and the Majorana phases gives an interference-limited floor
\begin{equation}
    |U_{\alpha N}|^2_{\rm min} = \frac{\big|\,m_a |U_{\alpha a}|^2 - m_b |U_{\alpha b}|^2\,\big|}{m_N}\,,
    \label{eq:flavor_floor}
\end{equation}
which is nonzero for generic PMNS parameters but can be much smaller than $\sum_i m_i/m_N$~\cite{Krasnov:2023jlt}.
The flavor pattern follows from the PMNS structure~\cite{Drewes:2016jae}: in NO the heaviest state $\nu_3$ dominates and its electron content is $|U_{e3}|^2=\sin^2\theta_{13}$, so $|U_{eN}|^2$ is strongly suppressible (asymptotically $|U_{eN}|^2/U^2 \gtrsim 0.5\%$ for $\eta\gg1$), while $|U_{\mu N}|^2 \simeq |U_{\tau N}|^2$; in IO both massive states carry $\mathcal{O}(1)$ electron content, so $|U_{eN}|^2$ is bounded from below ($|U_{eN}|^2_{\rm min}\simeq 0.019~\text{eV}/m_N$, about $19\%$ of the total floor) and $\mu,\tau$ become the suppressible directions.
For the electron flavor \cref{eq:flavor_floor} agrees with $\min_{\rm phases} m_{ee}/m_N$, where $m_{ee}$ is the effective Majorana mass probed in neutrinoless double-beta decay~\cite{Pascoli:2007qh,Krasnov:2023jlt}: the nonvanishing lower bound on $m_{ee}$ in IO is the same statement as the IO floor on $|U_{eN}|^2$, and with $m_1=0$ the NO electron mixing likewise cannot vanish.
The Majorana phase redistributes the mixing among flavors without changing the total $U^2$ in \cref{eq:U2_CI}; for $n=2$ only one combination is physical, $\alpha_{31}-\alpha_{21}$ in NO and $\alpha_{21}$ in IO. The real part $\xi$ drops out of \cref{eq:U2_CI,eq:Ualpha2}, but it controls how the mixing is shared between $N_2$ and $N_3$ and therefore enters the asymmetry.

\subsection{Baryon asymmetry computation}

For given values of $m_N$, $U^2$, the mass splitting $\delta M \equiv M_3 - M_2$, and the phases $(\xi, \alpha_{21}, \alpha_{31})$, the baryon-to-photon ratio $\eta_B$ is obtained by solving the momentum-averaged quantum kinetic equations (QKEs) of Refs.~\cite{Drewes:2017zyw,Hernandez:2015wna,Ghiglieri:2017gjz} for the $2\times 2$ HNL density matrix $\rho_N$ and the flavored lepton charges $\Delta_\alpha$ ($\alpha = e,\mu,\tau$), as functions of $z = T_{\rm sph}/T$, from $z=10^{-6}$ to $z=1$, starting from vanishing HNL abundances and lepton charges.
These equations include the helicity-conserving and helicity-flipping production and destruction rates~\cite{Ghiglieri:2017gjz}, the CP-violating oscillations of $\rho_N$ driven by $\delta M$, and spectator effects encoded in the susceptibility matrix relating chemical potentials to the conserved charges~\cite{Drewes:2017zyw}.
We integrate them with the public code \texttt{ULYSSES}~\cite{Granelli:2020pim,Granelli:2023knf}.
We use its \texttt{BEARS} module, which employs momentum-averaged, temperature-independent rates in the relativistic limit; once the HNL oscillations are fast, the off-diagonal elements of $\rho_N$ are averaged out.

To delineate the parameter space compatible with the observed asymmetry $\eta_B^{\rm obs} = 6.1 \times 10^{-10}$~\cite{Planck:2018vyg}, we perform a scan over the free parameters of the Casas--Ibarra matrix.
For each HNL mass $m_N$ in a grid of logarithmically spaced points from $0.1$ to $100~\gev$, we vary $U^2$ (equivalently $\eta$) across a grid spanning from the seesaw floor to $U^2 = 10^{-4}$, with bisection refinement to locate the boundaries of the contiguous viable interval.
At each $(m_N, U^2)$ point we scan the relative splitting $\log_{10}(\delta M/m_N)$ over three decades on either side of $\delta M/m_N \sim T_{\rm sph} H(T_{\rm sph})/m_N^2$, the splitting for which vacuum oscillations become effective near sphaleron freeze-out, and over combinations of the CP-violating phases $(\xi, \alpha_{21}, \alpha_{31})$ at the NuFIT best-fit Dirac phase, $\delta_{\rm CP}=197^\circ$ (NO) and $286^\circ$ (IO).
A point is deemed viable if any phase and splitting combination yields $|\eta_B| \geq \eta_B^{\rm obs}$.
The per-flavor projections $|U_{\alpha N}|^2$ are then obtained from \cref{eq:Ualpha2} by scanning both signs of $\eta$ and the Majorana phases within the viable $U^2$ interval at each mass; the resulting per-flavor edges are therefore the envelope compatible with successful leptogenesis, evaluated at fixed $\delta_{\rm CP}$.
We have verified at representative masses that the per-flavor lower edges are attained by parameter points with $|\eta_B|\geq\eta_B^{\rm obs}$.

\subsection{Results}

\begin{figure*}
    \centering
    \includegraphics[width=\textwidth]{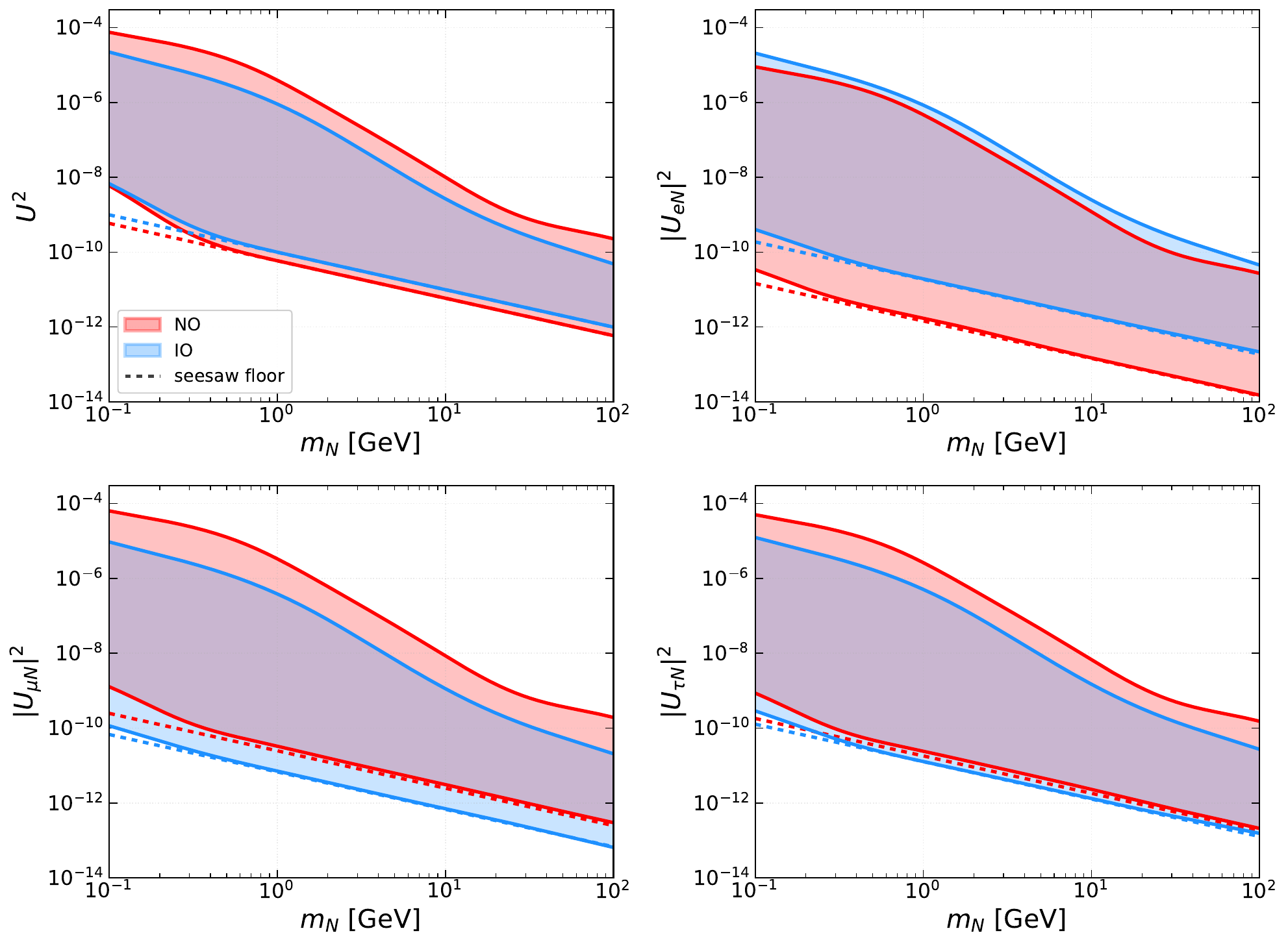}
    \caption{Leptogenesis target regions in the $(m_N,\,|U_{\alpha N}|^2)$ plane for the
    minimal $n=2$ type-I seesaw, for normal (red) and inverted (blue) mass ordering.
    The upper-left panel shows the total mixing;
    the remaining panels show the individual flavor mixings $|U_{\alpha N}|^2$ for
    $\alpha=e,\mu,\tau$. Shaded regions reproduce the observed baryon asymmetry
    $\eta_B^{\rm obs}$.
    Dotted lines show the corresponding analytical seesaw floors: $\sum_i m_i/m_N$ for the total
    mixing, \cref{eq:seesaw_floor}, and the interference-limited floor
    \cref{eq:flavor_floor} for each flavor. The lower edge of the total band approaches its floor around $m_N\simeq0.5~\gev$ and tracks it above $1~\gev$, as do the per-flavor edges above $1~\gev$, so the lower boundary of the target is
    set by the light neutrino masses rather than by leptogenesis; at lower masses a larger mixing is required, since at fixed $U^2$ the Yukawa coupling $Y^2=U^2m_N^2/v^2$ becomes too small to produce the HNLs before sphaleron
    freeze-out. Oscillation
    parameters are fixed to the NuFIT~5.3 best fit, including $\delta_{\rm CP}$.}
    \label{fig:leptogenesis_bands}
\end{figure*}
The resulting target bands are shown in \cref{fig:leptogenesis_bands}.
For both orderings a viable band exists across the full mass range $0.1$--$100~\gev$, spanning roughly 4--5 orders of magnitude in $U^2$ at $m_N \sim 1~\gev$ and narrowing at higher masses, because the washout boundary decreases more rapidly with mass than the seesaw floor.
The IO seesaw floor is higher than the NO floor by a factor $\simeq 1.7$, \cref{eq:seesaw_floor}, and the NO band extends to larger mixing than the IO band at its upper (washout) edge, by a factor $\sim 3$--$5$.
A plausible interpretation is that in NO the electron direction is only weakly coupled ($|U_{eN}|^2\propto\sin^2\theta_{13}$), so a lepton asymmetry stored there survives washout up to larger total coupling, whereas in IO the washout is more evenly distributed among flavors over most of the phase space; the precise factor is, however, limited by the resolution of our phase and $\delta M$ grids.

The lower edge of the summed band approaches the seesaw line, \cref{eq:seesaw_floor}, around $m_N\simeq0.5~\gev$ and tracks it closely above $1~\gev$, for both orderings: minimal-coupling seesaw HNLs already generate a sufficient asymmetry there, so the lower boundary of the target region is set by neutrino masses rather than by leptogenesis.
At lower masses the lower edge rises above the seesaw line.
This follows from the relation between mixing and Yukawa coupling, $Y^2 = U^2 m_N^2/v^2$: at fixed $U^2$ a lighter HNL has a smaller Yukawa and is produced more slowly, $\Gamma_N\sim Y^2 T$, so that too few HNLs exist by sphaleron freeze-out to carry the asymmetry.
Successful leptogenesis then requires $U^2 \gtrsim v^2 Y^2_{\rm min}/m_N^2\propto m_N^{-2}$, which overtakes the seesaw scaling $\propto m_N^{-1}$ at low mass and requires $|\eta|>0$, enhancing the mixing above the seesaw value.
Conversely, the upper edge reflects the washout of the asymmetry at large mixing, where the HNLs equilibrate before sphalerons decouple.

For the per-flavor bands, the lower edges coincide with the interference floor of \cref{eq:flavor_floor} for $m_N\gtrsim 1~\gev$, within $\sim 10\%$ for IO and $\sim 30\%$ for NO, and are lifted above it at low mass by the same production-rate argument.
The per-flavor lower edges are thus, like the summed one, a consequence of the seesaw structure rather than of leptogenesis over most of the mass range.
The electron mixing in NO can be suppressed to $\lesssim 1\%$ of the total, while in IO it is bounded from below at $\gtrsim 19\%$ of the total floor and $\mu,\tau$ become the suppressible flavors.
All bands are computed at the NuFIT best-fit Dirac phase; since $\delta_{\rm CP}$ remains poorly constrained, we note that only the IO $\mu$ and $\tau$ lower edges are sensitive to this choice, and would decrease by a factor $\sim 3$--$5$ if $\delta_{\rm CP}$ were varied over its allowed range, whereas the total seesaw floor is independent of $\delta_{\rm CP}$ and the electron and NO floors depend on it only weakly; we have not varied $\delta_{\rm CP}$ in the baryogenesis calculation itself.

We show the bands only for $m_N \geq 0.1~\gev$.
Although the asymmetry itself is generated before sphaleron freeze-out, well before BBN, HNLs with
mixing close to the seesaw line are long-lived, $\tau_N\propto(G_F^2U^2m_N^5)^{-1}$, and can decay during or after BBN, where their decays, mainly into active neutrinos, raise the neutron-to-proton ratio and hence the primordial $^4$He abundance~\cite{Dolgov:2000jw,Sabti:2020yrt}.
This is the same constraint already shown as the gray hatched region in
\cref{fig:muon_decay_sensitivity,fig:balloon_sens}, which excludes the small-mixing region
at low $m_N$ where the seesaw floor lies.
The bands shown impose the baryogenesis requirement but not BBN constraints, which can exclude part of the low-mass region, including masses above $0.1~\gev$.
Where they overlap, our bands are in good agreement with those of Refs.~\cite{Drewes:2016jae,Drewes:2017zyw,Klaric:2021cpi}.

\section{Conclusions} \label{sec:combined}
\label{sec:conclusion}

We have examined the capability of a next-generation muon facility as a source of heavy neutral leptons.
The relevant property of such a facility is the simultaneous availability of a multi-\tev\ beam energy, which keeps deep-inelastic HNL production kinematically open up to $m_N \sim \sqrt{2 m_p E_\mu} \approx 100~\gev$, and an integrated intensity of $N_\mu \sim 10^{22}$ muons, comparable to that of dedicated proton beam dumps.
Three experimental proposals exploiting this combination have been developed: a heavy neutrino factory, a muon beam dump with three target--decay-region configurations, and a muon-ion collider.
Their production mechanisms, geometries, backgrounds, and projected sensitivities are summarized below.

In the heavy neutrino factory of \cref{sec:decay}, HNLs are produced in $\mu \to N e \nu$ in a storage ring and observed through $N \to \nu e^+ e^-$ in a low-density detector of length $L_{\rm det} = 100~{\rm m}$ and radius $R_{\rm det} = 5~{\rm m}$ aligned with a straight section.
The differential production spectrum is given by \cref{eq:mu_diff_width} and the visible partial width by \cref{eq:Nee_width}; the geometric acceptance $\epsilon_{\rm det}$, evaluated by Monte Carlo, increases with both $R_{\rm det}$ and $E_\mu$.
Under the assumption of a background-free measurement, the resulting reach in $\UNsq{\mu}$ exceeds the LSND constraint of Ref.~\cite{Ema:2023buz} by approximately two orders of magnitude over the kinematically accessible interval $2 m_e < m_N < m_\mu$.
While this mass regime is already constrained by BBN~\cite{Sabti:2020yrt}, the heavy neutrino factory is also sensitive to non-minimal HNL models that evade cosmological bounds.
As an example, the sensitivity to HNLs that additionally couple through a transition magnetic moment is shown in \cref{fig:rates_dipole}.

In the dense-target beam-dump configuration of \cref{sec:dump}, HNLs are produced by deep-inelastic upscattering $\mu^\pm I \to N X$ in a dense target and decay in a downstream detector.
Two target benchmarks were considered, tungsten with $\ell_{\rm tar} = 40~{\rm m}$ and water with $\ell_{\rm tar} = 10~{\rm m}$, in both cases followed by a shield of $\ell_{\rm sh} = 10~{\rm m}$ and a decay region of $\ell_{\rm dec} = 100~{\rm m}$.
The signal yield is given by \cref{eq:Nsigdump}.
The dominant background arises from neutrinos produced by muon decays in flight and by muon--nucleus scattering in the target, which subsequently interact in the decay volume.
This decay region should be as low-density as possible, as the background rate is proportional to the density of the medium.
For a helium-filled volume at $1~{\rm atm}$ we obtain $B \approx 1.7$--$2.5 \times 10^{5}$ events for a $\mu^-$ beam on tungsten (\cref{tab:bgevents}).
Evacuating the decay volume to the pressure of \cref{eq:vacP}, which lies between the SHiP design specification and the operating vacuum of NA62, reduces the expected background below $0.1$ events and improves the reach in $\UNsq{\mu}$ by over an order of magnitude, as shown in \cref{fig:WsensDumpAll}.
Restricting the signal to final states with exactly two charged tracks forming a vertex that points back to the target somewhat reduces the sensitivity, as shown in \cref{fig:WsensDumpDilep}.
The water target is less sensitive than the tungsten target, as shown in \cref{fig:H2OsensDumpAll,fig:H2OsensDumpDilep}, but produces a smaller neutrino flux and relaxes the vacuum requirement by a factor of approximately six.

In the atmospheric beam-dump configuration of \cref{sec:balloon}, the beam is directed into the Earth at $2.3^\circ$ above the horizon from a depth of $100~{\rm m}$, traversing $2.4~{\rm km}$ of bedrock, and the resulting HNLs decay in the atmosphere.
Muon energy loss along the rock path was treated with \cref{eq:muon_range} and the accumulated multiple-Coulomb-scattering deflection with \cref{eq:muon_mcs}.
Production rates were computed with \cref{eq:terrestrial_HNL_production}, integrating the cross section over the degrading beam energy. 
The geometry benefits from the curvature of the Earth: a near-tangential trajectory reaches a detector altitude $h$ only after $d \approx \sqrt{2 R_\oplus h} \approx 250~{\rm km}$ for $h \sim 5~{\rm km}$, providing a decay baseline two orders of magnitude larger than the detector altitude. 
Cherenkov emission from the charged decay products is imaged by a TRINITY-style SiPM camera of radius $4~{\rm m}$ with $0.3^\circ$ pixels. 
The signal definition was taken to be HNL decays to dimuons, where the Cherenkov photons from each muon appear above threshold in two different pixels on the camera.
This definition removes the majority of backgrounds from neutrino scattering in the atmosphere; the only that remain are charged-current $\nu_\mu$ interactions creating a charm hadron that promptly decays to a muon, mimicking the dimuon HNL decay signature.
As shown in \cref{fig:dimuon_observables}, these neutrino-induced dimuon backgrounds can be distinguished from HNLs by rejecting visible hadronic showers and measuring the angle of the dimuon system with respect to the beamline.
\Cref{fig:balloon_sens} shows the resulting background-limited sensitivity of the atmospheric beam dump, which demonstrates similar reach to the dense-target configuration of \cref{sec:dump}.
If neutrino backgrounds can be removed completely, the $\UNsq{\mu}$ sensitivity extends down to the seesaw floor.

\Cref{sec:terrestrial_vacuum} presents the terrestrial-plus-vacuum beam-dump configuration, pairing the kilometer-scale rock target of the atmospheric setup of \cref{sec:balloon} with the vacuum decay region of \cref{sec:dump}.
This maximizes HNL production while suppressing backgrounds from neutrino scattering.
This setup has stricter pressure requirement in \cref{eq:terrestrial_pressure_req}, though above that demonstrated by NA62~\cite{NA62:2023qyn}.
This configuration has the best $\UNsq{\mu}$ reach among the configurations studied here and concretely reaches the seesaw floor.

In the muon-ion collider of \cref{sec:muon_ion}, the muon beam is brought into collision with a $7~\tev$ proton beam, giving $\sqrt{s} \approx 6.5~(11.8)~\tev$ for $E_\mu = 1.5~(5)~\tev$, and HNLs are reconstructed as displaced vertices in the fiducial range $L_0 = 0.4~{\rm mm} < d < L_1 = 11~{\rm m}$ according to \cref{eq:Ndec}.
Treating the search as background free and adopting unit reconstruction efficiency, the $2\sigma$ sensitivity extends out to HNL masses $m_N \sim 100~{\rm GeV}$, as shown in \cref{fig:MuICSens}.
For an integrated luminosity of $200~{\rm ab}^{-1}$, chosen to match the FCC-ee exposure, the sensitivity shown in \cref{fig:MuICFCC} becomes comparable to the FCC-ee displaced $\mu j j$ projection and approximately a factor of three above the deepest FCC-ee full-detector displaced-vertex projection.
The distinguishing feature of the collider proposal is its mass coverage rather than its depth: the larger center-of-mass energy sustains sensitivity to $m_N \sim 100~\gev$, beyond the kinematic reach of the proposed $e^+e^-$ machines.

\begin{figure}[h]
    \centering
    \includegraphics[width=\linewidth]{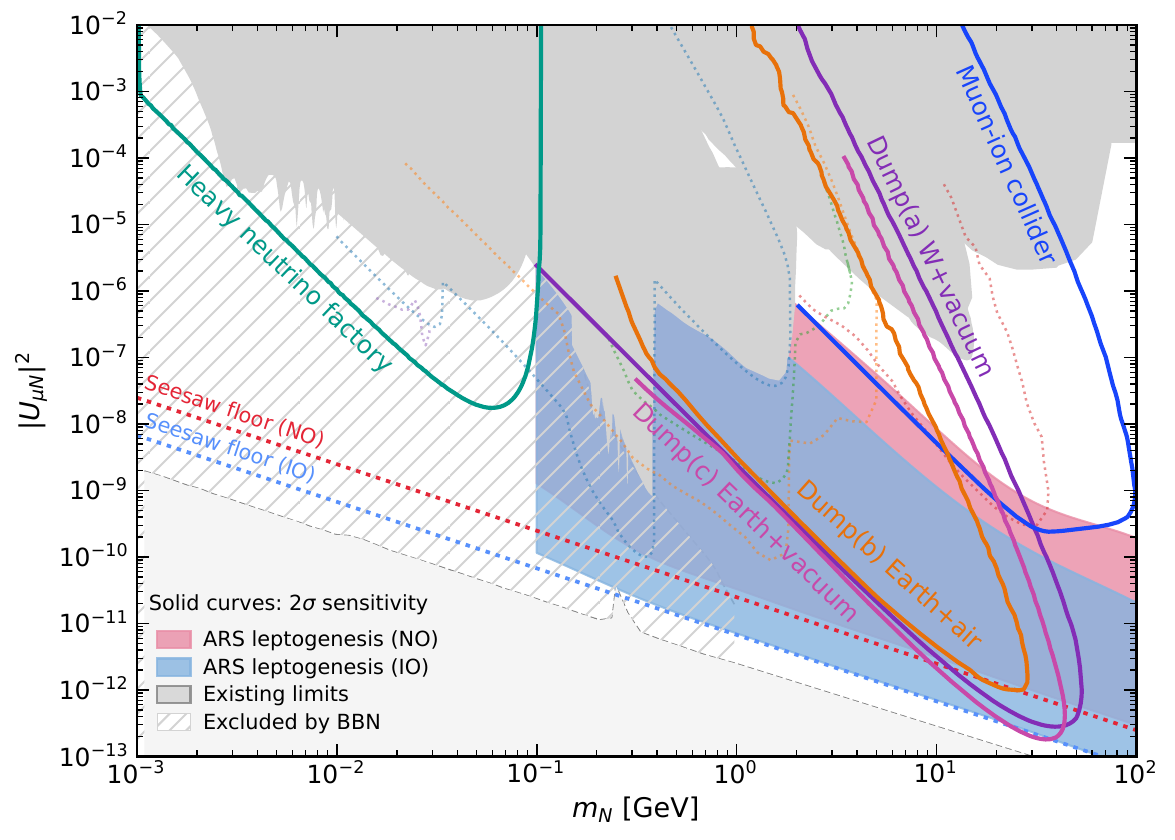}
    \caption{Combined $2\sigma$ projected sensitivity of the three proposals: the heavy neutrino factory, the muon beam dump in its three configurations, and the muon-ion collider. The projections are compared with existing constraints, BBN and with the parameter space compatible with the minimal type-I seesaw and ARS leptogenesis.
}
    \label{fig:combined}
\end{figure}

The combined result is presented in \cref{fig:combined}, which spans $10^{-3} \le m_N/\gev \le 10^{2}$.
Each solid contour delimits the region excluded at $2\sigma$ by one configuration for $N_\mu = 10^{22}$: green for the heavy neutrino factory of \cref{sec:decay}, which terminates at $m_N = m_\mu$; purple for the tungsten muon beam dump of \cref{sec:dump}, with an evacuated decay volume and the inclusive signal; orange for the atmospheric dump of \cref{sec:balloon} in its background-limited dimuon selection, which opens at the $\mu \mu$ threshold; magenta for the terrestrial-plus-vacuum beam dump of \cref{sec:terrestrial_vacuum}; and blue for the muon-ion collider of \cref{sec:muon_ion}, which begins at the assumed trigger threshold $m_N \gtrsim 2~\gev$ and remains open at the upper edge of the mass range shown.
Gray regions denote existing experimental constraints~\cite{Fernandez-Martinez:2023phj} and dotted curves the projections of other proposed experiments.
The hatched region is excluded by BBN~\cite{Sabti:2020yrt}, the red and blue dotted lines are the flavor-dependent seesaw floor of \cref{eq:flavor_floor} for normal and inverted ordering, and the pink and blue bands are the corresponding regions compatible with ARS leptogenesis.

Three features of \cref{fig:combined} are relevant for the interpretation of these results.
First, the five contours overlap pairwise, so that the combined coverage is continuous over $2 m_e \lesssim m_N \lesssim 100~\gev$, and over most of this interval the contours lie several orders of magnitude below the existing constraints.
Second, the two theoretical targets are not independent. The type-I seesaw fixes the minimum total mixing to $U^2_{\rm min} = \sum_i m_i/M_N$, equal to $0.059~{\rm eV}/M_N$ for normal ordering and $0.10~{\rm eV}/M_N$ for inverted ordering, and the ARS band obtained in \cref{sec:leptogenesis} is bounded from below by this same quantity and from above by the washout boundary.
Third, the seesaw floor is crossed. The tungsten beam dump with an evacuated decay volume lies below the muon-flavor floor of \cref{eq:flavor_floor} for normal ordering over $m_N \approx 12$--$52~\gev$, by up to a factor of $2.5$, and remains a factor of $1.5$ above the floor for inverted ordering.
The terrestrial-plus-vacuum configuration extends to the floor over approximately $m_N \sim 20$--$30~\gev$.
To the best of our knowledge, no previously proposed experiment attains the type-I
seesaw floor for $m_N > 1~\gev$.

Several sources of uncertainty affect these projections. 
The beam dump decay region in \cref{sec:dump,sec:terrestrial_vacuum} presupposes that the specified vacuum can be maintained across a $100~{\rm m}$ decay volume and that impact-parameter pointing and vertex resolution are sufficient to reject the residual neutrino flux, as demonstrated by SHiP and NA62, at comparable or larger neutrino intensities.
The atmospheric estimate assumes that a $4~{\rm m}$ radius SiPM camera can be built with sufficient pointing resolution to separate HNL decays and neutrino backgrounds.
Furthermore, assumptions on the total Cherenkov yield and angular distribution of \tev hadronic showers have a direct impact on the ability to reject neutrino backgrounds.
More detailed studies of the neutrino background, including the evaluation of higher-altitude free-flying or satellite-borne platforms that could render the inclusive selection background free, are left to future work.
The muon-ion projection assumes unit displaced-vertex reconstruction efficiency, and a realistic efficiency would weaken it by an $\mathcal{O}(1)$ factor.
In each case the degradation is of order unity rather than of order a decade, since the signal yields are steep functions of $\UNsq{\mu}$ in the vicinity of the quoted contours.
Because the configurations are limited by different systematics --- vacuum quality, atmospheric Cherenkov yield, and detector efficiencies --- the combined reach is correspondingly less sensitive to any single assumption than the individual contours.

Finally, we note that the parameters governing these searches --- the length and composition of the dump, the aperture of the near-beam detector, the extraction angle of the beam into the Earth, and the configuration of the muon-ion interaction region --- are largely fixed during the design of the accelerator complex itself.
The sensitivities reported here are therefore contingent on these choices being made with long-lived-particle searches in view.
This article is intended to emphasize that HNL searches should be included in the physics case for next-generation muon accelerators at the design stage.

\section{Acknowledgments}

We thank the organizers of the Good $\nu$'s workshop at Pitt PACC, during which initial discussions for this project took place.
We are grateful to Jackapan Pairin for providing a graphic design of our experimental concept.
The work of KX at MSU is supported in part by the U.S. National Science Foundation under Grants No. PHY-2112829, PHY-2013791, and PHY-2310497.
The work of KX at TDLI/SJTU is also supported by the National Natural Science Foundation of China and by the State Key Laboratory of Dark Matter Physics.
The work of P.-C. L. is supported by National Natural Science Foundation of China under the Grants No.12405121.
The work of G.H. is supported by the Neutrino Theory Network Fellowship with contract number 726844. 
N.K. is supported by the National Science Foundation (NSF, CAREER Grant \#2239795) and the David \& Lucile Packard Foundation.
C.A.A. is supported by the Faculty of Arts and Sciences of Harvard University, the National Science Foundation (NSF, CAREER Grant \#2239795), the John Templeton Foundation (Grant \#63651), the Research Corporation for Science Advancement, and the David \& Lucile Packard Foundation.
N.K. and C.A.A. are supported by the National Science Foundation under Cooperative Agreement PHY-2019786 (The NSF AI Institute for Artificial Intelligence and Fundamental Interactions, http://iaifi.org/).
This work made use of \texttt{MadGraph5\_aMC@NLO}~\cite{Alwall:2014hca}, the CT18NNLO
parton distributions~\cite{Hou:2019efy}, and \texttt{ULYSSES}~\cite{Granelli:2020pim,Granelli:2023knf}.

\bibliography{bib}

\appendix

\section{Heavy Neutrino Factory Sensitivity to Electron Mixing} \label{sec:muon_decay_electron}

\begin{figure}
    \centering
    \includegraphics[width=\linewidth]{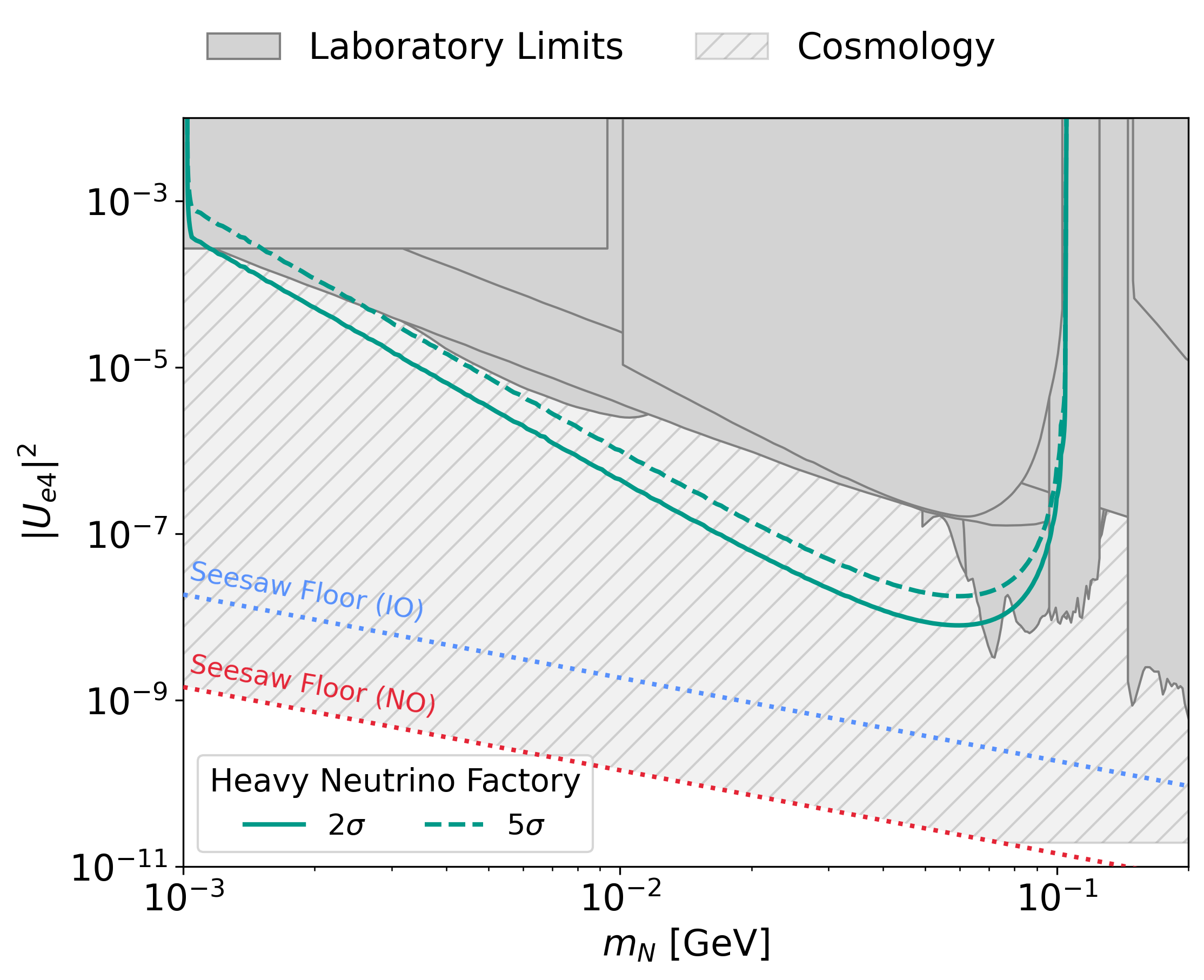}
    \caption{Sensitivity of the heavy neutrino factory setup to $|U_{eN}|^2$, in analogy to \cref{fig:muon_decay_sensitivity}.}
    \label{fig:muon_decay_electron}
\end{figure}

The heavy neutrino factory described in \cref{sec:decay} is also sensitive to electron-flavored mixing $|U_{e N}|^2$.
Production occurs by replacing the $\overline{\nu}_e$ in the muon decay diagram of \cref{fig:mu_decays} with an HNL $N$, described by \cref{eq:mu_diff_width}.
Detection also occurs via the $N \to \nu e^+ e^-$ decay channel, described in \cref{eq:Nee_width}, which now includes contributions from both the neutral current diagram of \cref{fig:mu_decays} as well as an additional a charged current contribution.
The resulting sensitivity of the heavy neutrino factory to $|U_{e N}|^2$ is shown in \cref{fig:muon_decay_electron}.

\section{Attenuation of Cherenkov photons in the atmosphere}
\label{sec:cherenkov_attenuation}

Cherenkov photons produced by relativistic muons propagating in the atmosphere are attenuated as they traverse the atmosphere before reaching a detector. In the near-UV and optical wavelength range relevant for Cherenkov detection ($\sim 300$-$600$~nm), attenuation is dominated by elastic scattering processes, while true absorption plays a subleading role.

We model atmospheric attenuation as an exponential loss,
\begin{equation}
T(\lambda) = \exp\!\left[-\tau_{\rm slant}(\lambda)\right],
\end{equation}
where $\tau_{\rm slant}$ is the slant optical depth accumulated along the photon trajectory. The vertical optical depth is written as the sum of molecular (Rayleigh) and aerosol (Mie) contributions,
\begin{equation}
\tau_{\rm vert}(\lambda) = \tau_{\rm R}(\lambda) + \tau_{\rm A}(\lambda).
\end{equation}
For Rayleigh scattering we use an empirical parameterization for the vertical optical depth from ground to space \cite{1995ApOpt..34.2765B},
\begin{equation}
\tau_{\rm R}^{\rm ground}(\lambda)
\simeq 0.00879 \left(\frac{\lambda}{\mu{\rm m}}\right)^{-4.09},
\end{equation}
which reproduces standard atmospheric transmission models in the optical band and is widely adopted in Cherenkov and atmospheric optics calculations . The vertical distribution of molecular scatterers is assumed to follow an exponential profile with scale height $H_{\rm mol}\simeq 8$~km, consistent with the U.S. Standard Atmosphere \cite{1992P&SS...40..553N}.

Aerosol attenuation is parameterized in terms of an aerosol optical depth (AOD) at a reference wavelength of 500~nm \cite{2012auger, auger2013ICRC...33.2322M},
\begin{equation}
\tau_{\rm A}^{\rm ground}(\lambda)
= \tau_{\rm A,500}
\left(\frac{\lambda}{500~{\rm nm}}\right)^{-\alpha},
\end{equation}
where $\alpha$ is the \AA ngstr\"om exponent, taken to be $\alpha\simeq 1$. Aerosols are modeled with a shorter exponential scale height $H_{\rm aer}\simeq 1$--2~km, reflecting their confinement to the lower troposphere. To bracket realistic observing conditions, we consider three representative aerosol scenarios: a \emph{clear} atmosphere with $\tau_{\rm A,500}=0.02$, a \emph{typical} atmosphere with $\tau_{\rm A,500}=0.05$, and a \emph{hazy} atmosphere with $\tau_{\rm A,500}=0.10$.

Both molecular and aerosol extinction coefficients decrease exponentially with altitude. The vertical optical depth above altitude $h$ satisfies
\begin{equation}
\tau_{\rm vert}(>h) = \tau_{\rm vert}^{\rm ground}\, e^{-h/H},
\end{equation}
with the appropriate scale height $H$. For optical wavelengths, more than $\sim 90\%$ of the molecular optical depth is accumulated below $\sim 30$~km altitude, while aerosol extinction becomes negligible above $\sim 10$--15~km \cite{Bernlohr:2008kv}. Cherenkov photon attenuation is then primarily determined by the emission altitude and propagation geometry, and is largely insensitive to detector altitude once the detector is located above the bulk of the atmosphere. This implies that balloon-borne detectors at $\sim 35$~km and space-based detectors at orbital altitudes should experience nearly identical atmospheric transmission.

For a photon propagating at zenith angle $\theta$ relative to the local vertical, we approximate the slant optical depth as
\begin{equation}
\tau_{\rm slant}(\lambda) \simeq X(\theta)\,\tau_{\rm vert}(\lambda),
\qquad
X(\theta)=\sec\theta,
\end{equation}
which captures the dominant increase in atmospheric column density for inclined trajectories and is valid for $\theta\lesssim 75^\circ$ \cite{Bernlohr:2008kv}. The zenith angle $\theta$ here refers to the photon propagation angle through the atmosphere and is not related to the intrinsic decay kinematics of the heavy neutral lepton, but is instead fixed by the geometric configuration of the decay point and the detector.

Cherenkov emission is weighted toward shorter wavelengths, with a differential spectrum $dN/d\lambda \propto \lambda^{-2}$. We therefore define a band-averaged transmission probability,
\begin{equation}
\langle T \rangle =
\frac{\int_{\lambda_{\rm min}}^{\lambda_{\rm max}}
T(\lambda)\,\lambda^{-2}\, d\lambda}
{\int_{\lambda_{\rm min}}^{\lambda_{\rm max}} \lambda^{-2}\, d\lambda},
\end{equation}
where we take $\lambda_{\rm min}=300$~nm and $\lambda_{\rm max}=600$~nm.

\begin{figure}
    \centering
    \includegraphics[width=\linewidth]{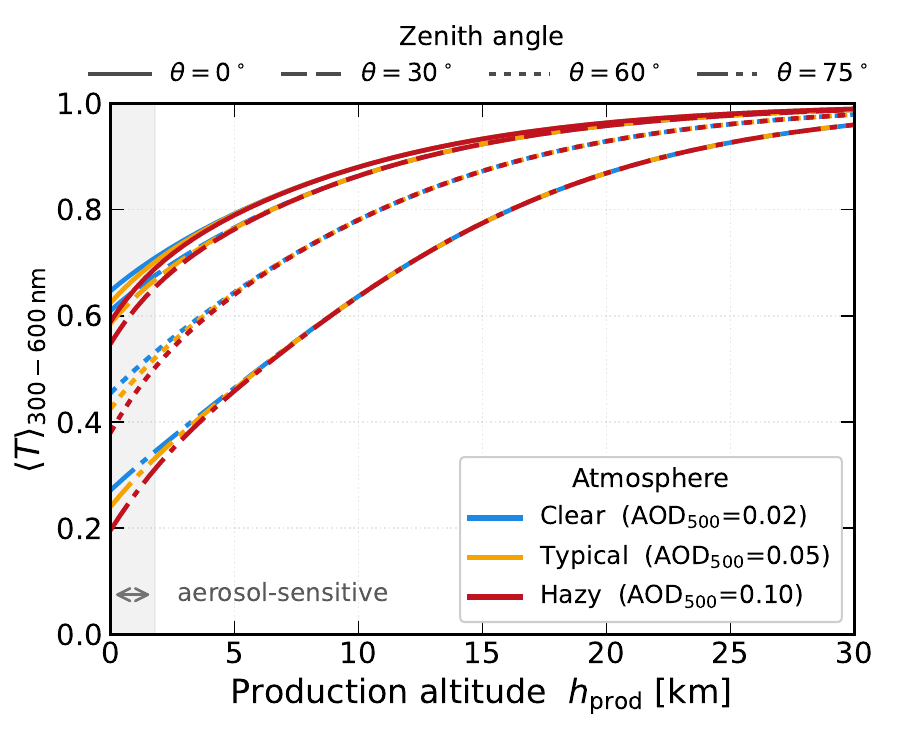}
    \includegraphics[width=\linewidth]{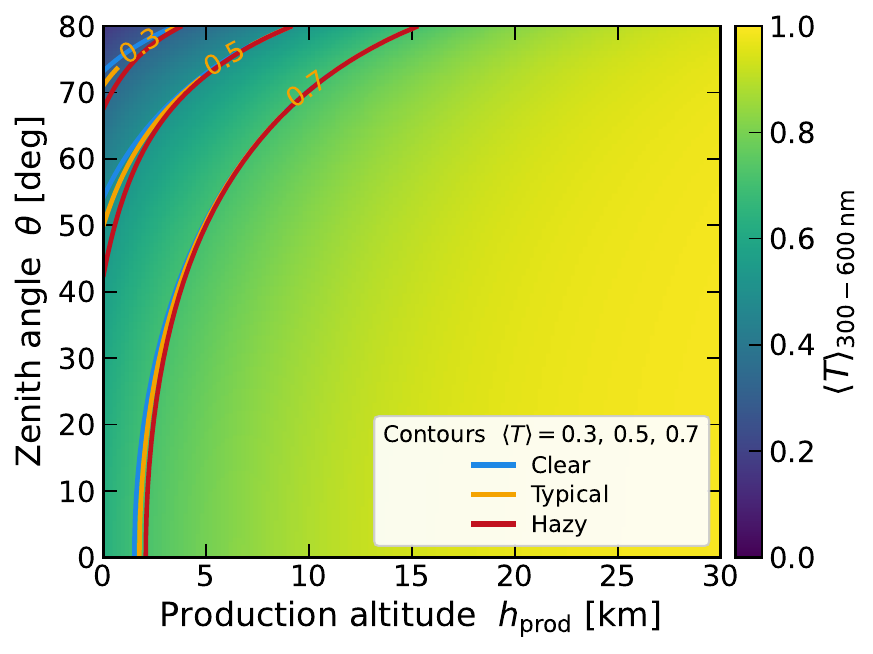}
    \caption{\textbf{Top:} The band-averaged Cherenkov photon survival probability as a function of production altitude, for three different atmospheric conditions.
    \textbf{Bottom:} The band-averaged survival probability as a function of zenith angle and production altitude, with contours indicating average transmissions probabilities of 0.3, 0.5 and 0.7 for three different atmospheric conditions (see main text).
    }
    \label{fig:attenuation_atmosphere}
\end{figure}

\end{document}